\documentclass[aps,pre,twocolumn,groupedaddress,showpacs]{revtex4}

\usepackage{latexsym,color,epsfig}

\usepackage{dcolumn}
\usepackage{bm}
\usepackage{amsmath}
\usepackage{amssymb}
\usepackage{amstext}
\usepackage{amscd}
\usepackage{setspace}
\usepackage{curves}
\usepackage{epic}

\def\aref#1{Appendix~\ref{#1}}
\def\eref#1{Eq.\thinspace\ref{#1}}
\def\erefs#1{Eqs.\thinspace\ref{#1}} 
\def\sref#1{Sect.~\ref{#1}}

\def\fref#1{Fig.~\ref{#1}}
\def\frefs#1{Fig.~\ref{#1}}
\def\Fref#1{Figure~\ref{#1}}

\newcommand{\av}[1]{\bigl\langle #1 \bigr\rangle}
\newcommand{\avtauinv}{\langle \tau^{-1} \rangle}

\def\kapf{\kappa_{\text{f}}}
\def\kaps{\kappa_{\text{s}}}
\def\kapt{\kappa}
\def\kttau{\kapt\tau_a}
\def\Ci#1{\av{f}^{#1}}

\def\Di#1{\mathcal{D}^{(\infty)}_{#1}}
\def\Cz#1{\av{f^{#1}}}
\def\Dz#1{\mathcal{D}^{(0)}_{#1}}
\def\kB{\text{k}_{\text{\tiny B}}}
\def\kT{\text{k}_{\text{\tiny B}}\text{T}}

\begin{document}

\title{Expanding the Temporal Analysis in Single-Molecule Switching
Experiments Through the Auto-Correlation Function: Mathematical
Framework}

\author{Darren E. Segall}
\email{desegall@caltech.edu} \affiliation{Department of Applied
Physics, California Institute of Technology, 1200 East California
Boulevard, Pasadena, California 91125-9500, USA}

\date{\today}

\pacs{02.50.Fz 82.37.Np 87.15.La 82.37.Gk}

\begin{abstract}
A method is presented that, when used in conjunction with single
molecule experimental techniques, allows for the extraction of rates
and mechanical properties of a biomolecule undergoing transitions
between mechanically distinct states.  This analysis enables the
exploration of systems where the lifetimes of survival are of order of
the intrinsic time constant of the experimental apparatus; permitting
the study of kinetic events whose transition rates are an order of
magnitude (or two) larger than those that can be studied with
traditional averaging methods.  Using current experimental techniques,
this allows for the study of biomolecules whose lifetime of survival
in a given state are as low as milliseconds down to microseconds.  The
relevant observable is the auto-correlation function of the
experimental probe that is attached to the biomolecule of interest.
Obtaining these auto-correlations functions is challenging and our
analysis is based upon a series solution.  Although the general
solution is expressed in terms of a series, two important limits are
studied, whose solutions are obtained exactly.  These limits
correspond to physically opposing limits: where transitions between
the mechanically distinct states of the biomolecule occur on either
much faster or much slower time scales than those governing the motion
of the experimental probe.  Not only do these limits correspond to
physically opposing bounds, but it is also proven that they correspond
to mathematical bounds to the general solution.  Motivated by the
derivation of these opposing bounds, two series solutions for the
general case are then presented.  Armed with these two general series
solutions that approach the exact solution from opposing points, we
then present an error analysis for truncating each series at arbitrary
order and obtain a range of parameters for which this method could be
applicable to the study of the two state biomolecular problem.
Finally, both series (up to third order) are expressed for the two
state problem when the system obeys Markov statistics.  These
solutions should be amiable to the analysis of experimental data,
expanding temporal analysis of data from single molecule experiments.
\end{abstract}

\maketitle

\section{Introduction}

The advent of experimental techniques such as atomic force microscopy
(AFM) and optical- and magnetic-trapping has made it possible to
manipulate individual biological
molecules~\cite{finzi95,smith96,mehta97,rief97a,rief97b,guilford97,oberhauser98,veigel98,carrionvazquez99,marszalek99,rief99,veigel99,meiners00,oberhauser01,liphardt01,marszalek02,schwaiger02,baker02,lia03}.
These methods provide complementary insights to those obtained using
the techniques of solution biochemistry.  In particular, they allow
for direct interpretation of specific processes occurring on
individual molecules which include the fluctuational character of real
macromolecular trajectories.

The single molecule experiments that have measured the kinetics of
structural transitions of macromolecules have accessed a wide variety
of processes.  Some of the most compelling examples include the
determination of rates, including those associated with the folding
and unfolding of various
proteins~\cite{rief97b,oberhauser98,marszalek99,rief99,oberhauser01}
and RNA~\cite{liphardt01}, structural transitions of
polysaccharides~\cite{marszalek02}, protein-induced loop formation in
DNA~\cite{finzi95,lia03} and the binding and unbinding of myosin to
actin filaments~\cite{mehta97,guilford97,veigel99,baker02}.  Despite
the diversity of these experiments, many have been limited to the
study of kinetics where the lifetimes of the macromolecular states of
interest are much larger than the intrinsic time constant of the
experimental probe.

\begin{figure}
\begin{center}
\centerline{\epsfxsize=3.2truein \epsfbox{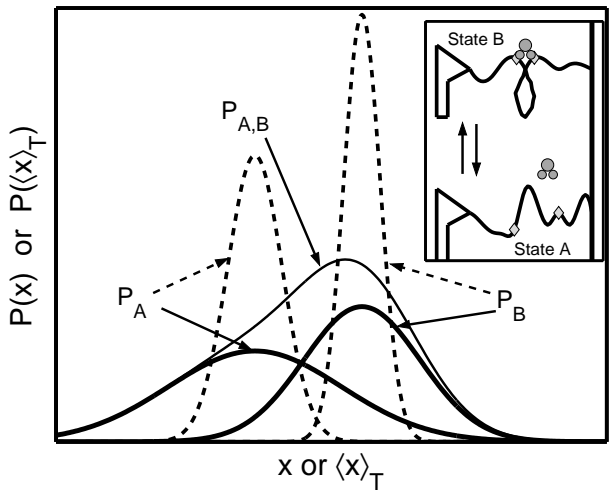}}
\caption{Inset: System of inspiration.  An AFM is attached to a
macromolecule which can exist in two states, an unlooped state (A) or
a looped state (B).  The looped state is caused by an interaction
between the macromolecule and a protein which binds to specific sites
(diamonds) on the macromolecule.  Figure: Different probability
distribution functions (P($x$)) of the displacement of the AFM tip
exists for each state (P$_\text{A}(x)$, P$_\text{B}(x)$: thick solid
lines).  For cases of interest the instantaneous distributions
(P($x$)) overlap.  If the position of the probe is averaged
($\langle$x$\rangle_{\rm T}$) over some time window T, then the
probability distribution function (P($\langle$x$\rangle_{\rm T}$):
dashed lines) of this averaged value will separate, provided T is
large enough and the likelihood of transitions during T is small.
P$_\text{A,B}$ (thin solid line): instantaneous joint probability
distribution function of the system.}
\label{fig:example}
\end{center}
\end{figure}

As an illustrative example of this aforementioned limitation and to
exemplify the types of problems that inspired this work we now
consider the abstract case where a macromolecule is attached to a
force probe (AFM, for example) and alternates between two mechanically
distinct states.  The inset in \fref{fig:example} displays the problem
of interest, where the macromolecule (DNA, for example) of concern
interacts with a protein that binds to specific sites (diamonds in
inset) on the macromolecule and loops out the intervening fragment.
The macromolecule then fluctuates between a looped and unlooped state.
(One can also consider folded/unfolded states or other types of
macromolecular interactions which cause conformational changes.)  From
the point of view of the experiment, these states can be distinguished
only if they give rise to different probability distributions for the
observable associated with the probe.  An example of this scenario is
displayed in the figure where the probability of finding the probe
displaced by a certain amount (P$_\text{A}$ or P$_\text{B}$: thick
solid lines in the figure) depends upon the state of the macromolecule
(state A or B).  (In the figure it has been assumed that the
distribution functions are Gaussian.)  Deleterious to the problem is
the fact that in order to have transitions between the macromolecular
states the probe must access positions where the molecule can exist in
either state.  Therefore, the distribution functions must overlap and
in many cases it is imperative that the distribution functions overlap
sufficiently in order that numerous transitions occur during an
experimental run.  Because of the requirement for overlap a single
measurement, in general, cannot determine the state of the system.  To
overcome this dilemma measurements $x$ ($x$:~displacement of the probe
for example) are averaged $\langle x\rangle_{\rm T}$ over a time
window T which reduces thermal noise.  If T is sufficiently large the
probability distribution functions of the averaged value
(P($\langle$x$\rangle_{\rm T}$): dash-lines in figure) will separate
and the averaged measurements ($\langle x\rangle_{\rm T}$) can discern
the states of the system.  (Alternatively, moments other than the
first ($\langle x\rangle_{\rm T}$) can be used to discern the states
of the system.  The arguments that follow here apply to such
measurements (see \sref{sec:background}).)

Given this framework we now illustrate the aforementioned limitation
that exists within single molecule experimental studies which, when
using such averaging procedures, requires the lifetimes of the states
to be much larger than the intrinsic time constant of the experimental
probe.  This limitation results from the fact that in order to obtain
well separated distribution functions (P($\langle$x$\rangle_{\rm T}$))
which properly correspond to the states of the system requires a
relationship between the time window T of averaging and the two
aforementioned time scales.  First, in order that the resulting
distribution functions P($\langle$x$\rangle_{\rm T}$) of the averaged
value $\langle x\rangle_{\rm T}$ correspond to a particular state
requires that the likelihood of a transition from one state to the
other is very low over the time period T.  This is guaranteed only if
the lifetimes of the states $\{\mathcal{T}_i\}$ (average survival
time) is much greater than the measuring period T, $\{\mathcal{T}_i\}
\gg$~T.  Secondly, in order to obtain well separated distributions
requires that sufficient thermal noise is reduced over the time window
T.  This is satisfied if the intrinsic time constant of the
experimental probe $\tau_{\rm P}$ is much smaller than the measuring
period T, $\tau_{\rm P} \ll$~T. (For an over-damped system $\tau_{\rm
P} = \gamma/k_{\rm P}$, where $\gamma$ is the damping constant of the
probe and $k_{\rm P}$ is its spring constant.)  The argument for this
relation goes as follows: Given an instantaneous measurement of the
probe $x_i$ which lies in the overlap region, for the averaged value
$\langle x\rangle_{\rm T}$ (for which this instantaneous measurements
$x_i$ contributes) to fall in one of two well separated distribution
functions requires the set of instantaneous measurements $\{x\}$
(which determines this averaged value $\langle x\rangle_{\rm T}$) to
represent a distinguishing characteristic of that particular
distribution function.  This requires the set of instantaneous
measurements $\{x\}$ to contain a range of values, implying it must
contain a number of uncorrelated measurements.  Roughly speaking, it
takes time of order of the intrinsic time constant of the experimental
probe $\tau_{\rm P}$ to make two uncorrelated measurement; therefore,
requiring $\tau_{\rm P} \ll$~T.  \Fref{fig:example} displays the
distribution functions of the averaged value,
P($\langle$x$\rangle_{\rm T}$), corresponding to a time window with
ten uncorrelated measurements, T $\approx 10~\tau_{\rm P}$.  Under
such conditions two ``separated'' distributions functions are
obtained.  The quantitative reduction of the width with respect to the
number of uncorrelated measurements is presented in \aref{app:altarg}.
(In principle, the time constant governing the motion of the probe
should include contributions from the molecule, however, for the
systems of interest the intrinsic time constant of the experimental
probe $\tau_{\rm P}$ gives the appropriate measure.  This point is
further elaborated on in \aref{app:tc})

These requirements have placed limitations on numerous experimental
studies that use single-molecule probe techniques: the lifetimes of
survival must be much larger than the intrinsic time constant of the
experimental probe $\{\mathcal{T}_i\} \gg \tau_{\rm P}$.  Optical trap
measurements~\cite{mehta97,guilford97,veigel98,veigel99,meiners00,liphardt01,baker02}
typically have a time constant of the order of 0.5~ms.  Such
techniques have been limited to the study of kinetic events where
states survive on times scales of the order of 10's of milliseconds
and, in general, much
longer.~\cite{mehta97,guilford97,veigel98,veigel99,liphardt01,baker02}

A number of biological molecules can fluctuate between different
states on time scales of the order of microseconds down to
nanoseconds.  Such measurements have been made using the techniques of
solution biochemistry~\cite{jacob99,mayor00,crane00,myers02}.  For a
single-molecule probe technique to have the capability of measuring
the properties of such fast switching systems requires the development
of either new experimental methods with higher temporal resolution or
alternative analytic methods capable of extracting properties of the
system from current experimental measurements.  New experimental
methods such as the placement of nanometer sized resonators on
standard AFM tips~\cite{hughes03,bargatin05} can provide a promising
future in the development of probes with both the high temporal
resolution and the large compliance needed to measure such properties.
However, such systems have not yet been fully developed for the study
of single biomolecular systems.

In this work we propose an alternate analytic approach to analyze data
from single-molecular probe techniques.  This analysis allows for the
exploration of macromolecular systems when the lifetimes of survival
of the states are on the order of the intrinsic time constant of the
experimental probe, $\{\mathcal{T}_i\} \approx \tau_{\rm P}$.  The
property computed is the position-position auto-correlation function
of the experimental probe.  The auto-correlation function is chosen
because it depends upon the joint probability distribution
P$_{\text{A,B}}$ (thin line in \fref{fig:example}) function and not on
the distribution function of the individual states (P$_\text{A}$ or
P$_\text{B}$).  This allows us to forgo the requirement that the
system remain in a particular state for a ``long'' period of time and
in turn increases the time resolution of the analysis.  (The
trade-off, however, is in the ability to relate the statistical
properties of the entire system (one which depends upon
P$_{\text{A,B}}$) to properties of the individual states (lifetimes
for example).)  Moreover, the auto-correlation function varies on time
scales of order of the intrinsic time constant of the experimental
probe; therefore, its form should be affected by events
(macromolecular transitions) that occur on such corresponding time
scales.  By appropriate analysis of the data, it should be possible to
determine lifetimes of states with characteristic scales ranging from
milliseconds to microseconds with current experimental techniques.

This manuscript focuses on the general mathematical formalism for
calculating the auto-correlation function of a system that is
switching between mechanically distinct states.  To further illustrate
the problems of interest we proceed in \sref{sec:background} with
numerical studies of a mock experiment (computer simulation)
representing the systems of interest.  These simulations reveal how
traditional methods (such as that described in this introduction) fail
to determine the rates of transitions of a biomolecule when
fluctuations occur on times scales of order of the intrinsic time
constant of the experimental probe.  In addition, presented are
numerical solutions for the auto-correlation function of the mock
system.  These numerical solutions illustrate how its form depends on
the value of the rates of transition and could be used as a conduit to
extract such rates from experimental measurements, those beyond the
reach of traditional analysis.

The following sections proceed with the mathematical formalism
associated with calculating the auto-correlation function.  Prior to
determining the general solution we focus on two special cases, termed
the fast-switching and slow-switching limits, \sref{sec:theory}.
These limits correspond to cases in which the macromolecule switches
between states on time scales that are either much faster or much
slower than the time scales associated with the motion of the
experimental probe.  In addition, ``closed''~\footnote{Closed form
solution is a vague definition (for example the exponential is general
considered closed form although it is really an infinite sum).  In
this article we refer to a closed form solution as one that can
written in terms of functions that can be readily evaluated.} form
solutions for auto-correlation function in these limiting scenarios
are presented and it is proven that these solutions correspond to
either a lower bound (fast-switching) or an upper bound
(slow-switching) to the general solution for the auto-correlation
function.

Although ``closed'' form solutions are found for these two special
cases, they do not exists in general, at least to our knowledge.
Therefore, solutions are provided for the auto-correlation function in
the form of a series solution, \sref{sec:expansion}.  Motivated by the
fact that the two bounds found for the general solution also
correspond to physically opposing limits, we generate two series
solutions for the auto-correlation function which initiated at these
opposing bounds.  Providing these two series solutions has the appeal
that they approach the exact solution from opposite points.

The convergence of these two series to the exact solution is examined
in \sref{sec:convergence}.  Additionally, this section addresses
issues regarding the applicability of this method to the analysis of
experimental data.  This section focuses mainly on the two state
problem and provides solutions for both series, truncated at third
order, for the two state Markov process.  These solutions should be
amiable to experimental analysis, where the rates of transition can be
determined in a regime inaccessible by traditional analysis.

\section{Illustrative Case: A Numerical Study}\label{sec:background}

\begin{figure}
\begin{center}
\centerline{\epsfxsize=3.0truein \epsfbox{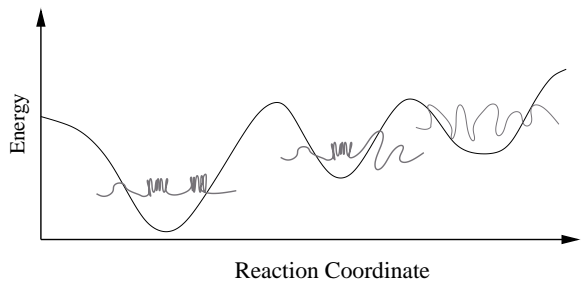}}
\caption{Energy landscape of a biomolecule undergoing structural
transformations between folded states. The reaction coordinate
corresponds to a physical degree of freedom along which the
biomolecule can make a transition from one state to the others.  Each
folded state corresponds to a well in the energy landscape and their
associated mechanical response are distinct.  Because of thermal
fluctuations the biomolecule can make transitions between the
different states.}
\label{fig:energylandscape}
\end{center}
\end{figure}

Our work is motivated primarily by biological molecules which can make
transitions between mechanically distinct states and the associated
single-molecule probe techniques used for detecting the properties of
these systems.  In general, the transitions we have in mind correspond
to structural transitions of the biomolecule which are induced either
by thermal fluctuations or imposed by interactions with other
macromolecules. (See inset in \fref{fig:example}, for example.)
Figure~\ref{fig:energylandscape} presents a schematic of the energy
landscape describing the conformational states of a biomolecule which
can exist in any of a number of different folded states.  (The energy
landscape may be thought of as the bare energy landscape of the
molecule or the resulting energy landscape which includes to the
presence of the experimental probe and is generally biased along a
particular reaction coordinate due to the stretching of the molecule.)
Each folded state has a different mechanical response and correspond
to distinct energy wells in the energy landscape of the biomolecule.
Because the biomolecule is undergoing thermal fluctuations it can make
transition between the various local wells.

\begin{figure}
\begin{center}
\centerline{\epsfxsize=3.2truein \epsfbox{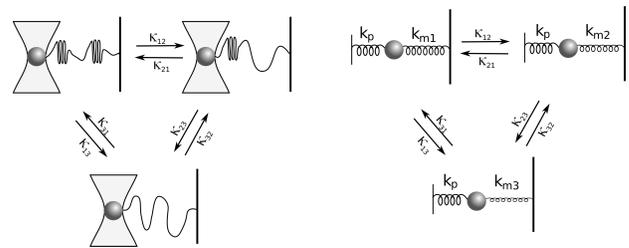}} 
\caption{An experimental scenario. Left: An optical bead in an optical
trap is connected to a biomolecule which can access multiple,
mechanically distinct states. The other end of the biomolecule is
anchored in place.  Transitions between the states are determined by a
set of rate constants ($\kappa_{ij}$).  Right: Schematic of
interactions in terms of springs. The probe (optical bead) has an
intrinsic spring constant ($k_{\text{P}}$) and each biomolecular state
has a specific spring constant ($k_{{\rm m}i}$).}
\label{fig:ob}
\end{center}
\end{figure}

In single molecule experiments like those being considered here, an
experimental probe is attached to the end of a biomolecule.  Each
detectable state differs in their mechanical response.  The
transitions between the states are governed by some set of rate
constants $\kappa_{ij}$, where the subscripts label the state being
exited and that being entered.  Figure~\ref{fig:ob} (left) displays a
typical example.  The experimental probe (optical bead in an optical
trap, for example) is held essentially at a fixed position (other than
motion induced by thermal noise) and the biomolecule is essentially
held at fixed extension.  The mechanical characteristics of the
biomolecular states will reveal themselves as mechanical springs whose
corresponding spring constants are determined by the state of the
system at that given extension~\cite{hatfield99}.  The plot on the
right displays a cartoon representing the mapping of the original
experimental scenario onto a corresponding mechanical model in terms
of coupled masses and springs.  The experimental apparatus and the
molecule in its different states have mechanical stiffnesses
characterized by spring constants $k_{\rm P}$ and $k_{{\rm m}i}$,
respectively, where $k_{{\rm m}i}$ differs for each molecular state
$i$.

As expressed in the introduction, the states of the system can be
determined, in general, only if the lifetimes of the states are
sufficiently large as compared to that of the intrinsic time constant
of the experimental probe, $\{\mathcal{T}_i\} \gg \tau_{\rm P}$.  To
give a more concrete representation of this limitation we now give a
specific study by way of numerical methods, where we perform mock
experiments to determine the feasibility of resolving the properties
of the different states of a macromolecule through methods similar to
that presented in the introduction.  Next, we present numerical
solutions for the auto-correlation function to demonstrate how an
analysis of the auto-correlation function can be used to alleviate
this limitation.  Having demonstrated the utility of analyzing the
auto-correlation we proceed this section with our theoretical
treatment for calculating the auto-correlation function.

\begin{figure}
\begin{center}
\setlength{\unitlength}{10mm}
\begin{picture}(6.5,4.6)(3,2)
 \linethickness{.3mm} \put(3,3){\vector(1,0){6.5}}
 \put(3,3){\vector(0,1){4.5}} \linethickness{1.pt}
 \dashline[+3]{.3}(3,3)(7,7)
 \put(4.5,6.5){\makebox(0,0)[t]{Under-Damped}}
 \put(8,6.5){\makebox(0,0)[t]{Over-Damped}}
 \put(6.,2.5){\makebox(0,0)[t]{$\left(\gamma/2\text{M}\right)^2
 \left[\frac{1}{\text{ms}^2}\right]$}}
 \put(2.25,5.5){\makebox(0,0)[r]{$\omega^2$}}
 \put(2.4,5){\makebox(0,0)[r]{$\left[\frac{1}{\text{ms}^2}\right]$}}
 \put(3,2.85){\makebox(0,0)[t]{10$^3$}} \put(3,2.9){\line(0,1){.2}}
 \put(4,2.85){\makebox(0,0)[t]{10$^4$}} \put(4,2.9){\line(0,1){.2}}
 \put(5,2.85){\makebox(0,0)[t]{10$^5$}} \put(5,2.9){\line(0,1){.2}}
 \put(6,2.85){\makebox(0,0)[t]{10$^6$}} \put(6,2.9){\line(0,1){.2}}
 \put(7,2.85){\makebox(0,0)[t]{10$^7$}} \put(7,2.9){\line(0,1){.2}}
 \put(8,2.85){\makebox(0,0)[t]{10$^8$}} \put(8,2.9){\line(0,1){.2}}
 \put(9,2.85){\makebox(0,0)[t]{10$^9$}} \put(9,2.9){\line(0,1){.2}}
 \put(2.7,3.3){\makebox(0,0)[t]{10$^3$}} \put(2.9,3){\line(1,0){.2}}
 \put(2.7,4.3){\makebox(0,0)[t]{10$^4$}} \put(2.9,4){\line(1,0){.2}}
 \put(2.7,5.3){\makebox(0,0)[t]{10$^5$}} \put(2.9,5){\line(1,0){.2}}
 \put(2.7,6.3){\makebox(0,0)[t]{10$^6$}} \put(2.9,6){\line(1,0){.2}}
 \put(2.7,7.3){\makebox(0,0)[t]{10$^7$}} \put(2.9,7){\line(1,0){.2}}
 \curvedashes[1mm]{0,1,2} \closecurve[1](5.5,3.3, 6.5,4.5, 7.8,5,
 9,4.5, 8,3.8, 7,3.5)
 \put(6,3.8633){\makebox(0,0)[t]{\scriptsize{\cite{liphardt01}}}}
 \put(7.9138,4.8451){\makebox(0,0)[t]{\scriptsize{\cite{guilford97,mehta97,baker02}}}}
\put(7.7076,4.4472){\makebox(0,0)[t]{\scriptsize{\cite{veigel98,veigel99,meiners00}}}}

 \put(8.5,3.5){\makebox(0,0)[l]{\footnotesize{OT}}}
 \linethickness{2.mm}
 \put(8.5,3.5){\vector(-3,1){.6}}

\end{picture}
\caption{Division between the over-damped and under-damped regimes as
a function of frequency $\omega$ and viscous damping $\gamma/2$M,
where $\gamma$ is the damping constant and M is the associated mass of
the probe.  Region where optical trap (OT) experiments lie is encased
in the dashed-circle.  The dashed line is the separation between the
over-damped and under-damped regions.}
\label{fig:OD}
\end{center}
\end{figure}
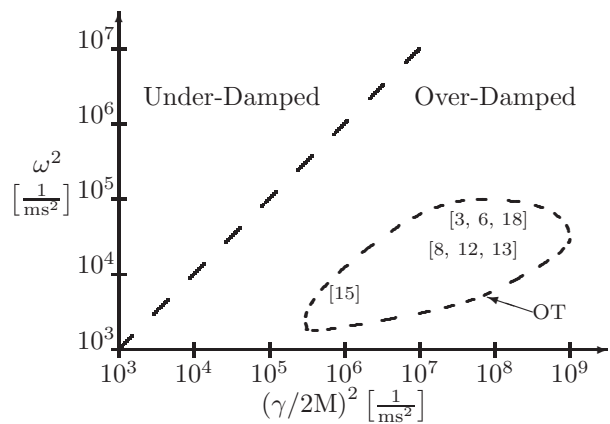

Prior to proceeding with this illustration expressed are a few key
assumptions imposed on the system, which in turn affects the form of
the equation of motion employed for the experimental probe.  These
assumptions are physically motivated and will be used throughout the
remainder of the paper.  First, it is assumed that the different
states of the biomolecule only affect the spring constant of the
biomolecule and do not alter the damping constant of the experimental
probe.  This assumption is justified on the grounds that the
experimental probe is, in general, much larger than the molecule of
interest and consequently, any fluid flow generated by the biomolecule
(or more importantly the differences in fluid flow from different
states) will only weakly influence the experimental probe.  As a
result, such differences are ignored.  Additionally, it is assumed
that the experimental probe is over-damped.  For most single molecule
experiments of interest this will be the case.  Figure~\ref{fig:OD}
charts the dividing line between the over-damped and under-damped
regimes for optical trap measurements, as a function of frequency of
the probe ($\omega$) and its viscous damping ($\gamma/2\text{M}$),
where $\gamma$ is the damping constant associated with the motion of
the probe and M is its mass.  Clearly these measurements are performed
in the over-damped limit.  Furthermore, it is assumed that once a
state (in one energy well) begins a transition to another state (to
another energy well) it occurs instantaneously or more specifically,
faster than all other time scales in the problem; allowing the
mechanical properties of the macromolecule at any given instant to be
described by a state corresponding to one of the energy wells.  In
addition, it is assumed that the interaction between the heat bath and
experimental probe can be expressed as a white, Gaussian noise
process.  Finally, in this manuscript it is assumed that ergodicity
holds, time averages are equal to ensemble averages.

\subsection{Limitations in averaging procedures}\label{sec:limitations}

We now present our numerical illustration.  Given the above
assumptions, the equation of motion for the probe is written using a
Langevin approach, in which the force acting on the probe exhibits
switching behavior characterized by different spring constants. It is
mathematically convenient to think of this problem as a single
Langevin equation with a constant damping constant, but with a
fluctuating spring constant and equilibrium position.  Expressed is
the equation of motion for the tethered probe:
\begin{equation}
\gamma \dot x(t) =  F_r(t) - k(t) (x(t) - x^o(t)). \label{eqn:case1}
\end{equation}
Here $x(t)$ is the position of the probe and the associated velocity
is given by $\dot x(t)$.  The fluctuating spring constant is
represented by $k(t)$, whose value takes $k_i$ when the system is in
state $i$.  The value for the fluctuating spring constant, $k_i$,
includes both the spring constant of the biomolecule and the intrinsic
spring constant of the experimental probe  
\begin{equation}
k_i = k_{\rm P} + k_{{\rm m}i}.\label{eqn:ki}
\end{equation}
The equilibrium position of the system $x^o(t)$ is also a fluctuating
variable, whose value is denoted as $x_i^o$ when the system is in
state $i$.  The damping constant, $\gamma$, is the same for each
state~\footnote{$\gamma$ can be considered the damping constant of the
experimental probe itself or the average damping constant including
influences from the surfaces and coupling to the biomolecule.}.
Finally, $F_r(t)$ is the random thermal force due to the presence of
the environment and is assumed to be Gaussian white noise
characterized by the correlations
\begin{eqnarray}
\av{F_r(t)} &=& 0,  \label{eqn:fm} \\
\av{F_r(0)F_r(t)} &=& 2\kT\gamma \delta(t), \label{eqn:sm}
\end{eqnarray}
where $\kB$ is the Boltzmann constant and T is the temperature.

To develop intuition for this problem, we choose the spring constant
for the experimental probe $k_{\rm P}$ to be 20~pN/$\mu$m and its
damping constant $\gamma$ to be 9.5 fg/ns.  Such values are typical of
optical beads in a laser trap~\cite{meiners00}.  The corresponding
intrinsic time constant ($\tau_{\rm P} = \gamma/k_{\rm P}$) of the
experimental probe equals 0.48 milliseconds.  We consider a
biomolecule which can exist in two states with corresponding spring
constants 25~pN/$\mu$m (state 1) and 5~pN/$\mu$m (state 2).  Spring
constants which are easily obtainable in muscle
proteins~\cite{rief97b,oberhauser98,marszalek99,rief99,oberhauser01}.
For simplicity, the equilibrium position of each state is chosen to be
equal, corresponding to observation of transverse motion of the
experimental probe. (Transverse to the extension of the biomolecule.)
The rate for the biomolecule to make a transition from state $i$ to
state $j$ is given by the rate constant $\kappa_{ij}$.

\begin{figure}
\begin{center}
\centerline{\epsfxsize=3.2truein \epsfbox{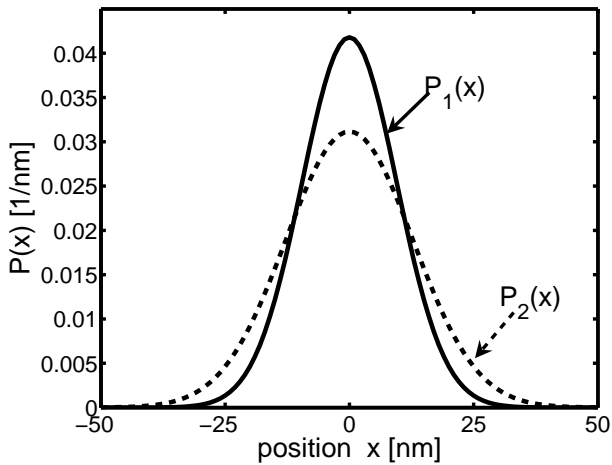}} 
\caption{The normalized distribution functions of the position of the
experimental probe when the molecule is in either state 1 or state 2.  The
spring constant of each state ($k = k_{\rm P} + k_{{\rm m}i}$) is equal to the
spring constant of the experimental probe ($k_{\rm P}$) plus the spring
constant of the molecule for that particular state ($k_{{\rm m}i}$).  For
state 1 the molecular spring constant is 25~pN/$\mu$m and for state 2
it corresponds to 5~pN/$\mu$m.  The spring constant of the
experimental probe is taken to be
20~pN/$\mu$m.}\label{fig:distribution}
\end{center}
\end{figure}

Before embarking on our numerical test we investigate properties of
the distribution function for the position of the experimental probe
when the molecule is in a given state.  These distribution functions
($\rho_i(x)$) are determined through Boltzmann statistics, each with
form
$$\rho_i(x) = \frac{e^{-\beta k_i x^2/2}}{Z_i}.$$ Here $\beta =
1/\kB$T is proportional to the inverse temperature and $Z_i$ is the
partition function (normalization) for state $i$. (Recall $k_i$ is the
spring constant for state $i$, \eref{eqn:ki}.)
Figure~\ref{fig:distribution} displays the two distribution functions.
Unlike the example in the introduction, here the average position of
the probe $\langle x\rangle_i$ (first moment) for each state cannot be
used to decipher the state of the system as they are equal, $\langle
x\rangle_1 = \langle x\rangle_2$.  In such cases the squared
displacement of the probe $\langle x^2\rangle_i$ (second moment)
differ for each state (due to differences in spring constants) and can
be used to decipher the state of the system.  The squared displacement
corresponds to a measurement of the width of a distribution function
squared.  Such procedures have been employed in single molecule
experiments~\cite{patlak93,mehta97,guilford97,veigel99,baker02}.

Although the measure used here differs from that proposed in the
introduction, the arguments from the introduction are still applicable
in revealing constraints on various time scales.  To obtain an
accurate measurement for the width of a distribution function requires
the time window of averaging T to be sufficient, such that a number of
uncorrelated measurements are made within this time window; requiring
the time window to be much greater than the intrinsic time constant of
the experimental probe, T$\gg \tau_{\rm P}$.  To obtain accurate
accounts for the individual widths of the distribution functions
requires the system to remain in a given state over this time window;
requiring the time window to be much less than the lifetimes of the
states, T$\ll \{\mathcal{T}_i\}$.  These two conditions lead to the
requirement that for this accounting method to be useful for exploring
the states of the system the intrinsic time constant of the
experimental probe must be much less than the lifetimes of the states,
$\tau_{\rm P} \ll \{\mathcal{T}_i\}$.

\begin{table}
\caption{\label{tab:rates} Rates of transition between the two
different states.  $\kappa_{ij}$ is the rate at which transitions are
made from state $i$ to state $j$.  The rates correspond to a range of
illustrative cases, which differs by powers of ten.  The rates are in
units of Hertz and in the parenthesis are the lifetimes of the states
(inverse of the rates) in units of milliseconds.}
\begin{ruledtabular}
\begin{tabular}{c|c|c|c|c}
rates & case A & case B & case C & case D \\
\hline
$\kappa_{12}$ [Hz] (ms) & 0.26 (3800) & 2.6 (380) & 26 (38) & 2600 (0.38) \\
$\kappa_{21}$ [Hz] (ms) & 0.11 (9100) & 1.1 (910) & 11 (91) & 1100  (0.91)
\end{tabular}
\end{ruledtabular}
\end{table}

Having illustrated that the time constraints presented in the
introduction still pertain to our numerical study, we return to our
numerical test and illustrate the limitations of a traditional
analysis for identifying the states of the macromolecule and
determining their associated properties.  To explore the applicability
of such traditional methods we considered a range of examples.  These
examples range with respect to lifetime of the states, from those
which are much greater than the intrinsic time constant of the
experimental probe ($\{\mathcal{T}_i\} \gg \tau_{\rm P} = 0.48$~ms) to
those which are of the same order as the intrinsic time constant of
the experimental probe ($\{\mathcal{T}_i\} \approx \tau_{\rm P}$).
Four cases are considered within this range, whose rates are displayed
in Table~\ref{tab:rates}.  The lifetimes of the states are displayed
in parentheses in the table and cover four orders of magnitude.

To obtain a numerical solution for the position of the probe, $x(t)$,
\eref{eqn:case1} is integrated using a standard Langevin dynamics
algorithm~\cite{vangunsteren82}~\footnote{Note that the algorithm used
here \cite{vangunsteren82} also includes an inertial term.  So long as
the mass is taken small, such that the system is over-damped
(\fref{fig:OD}), this algorithm produces the correct trajectories
governed by \eref{eqn:case1}}.  During the calculation the state of
the system and hence the spring constant ($k(t)$) is determined as
follows: If at time $t$ the system is in state $i$ then the
probability of the system at time $t + dt$ to be in state $j$ is
$\kappa_{ij} dt$.  Transitions are then determined by a uniform random
number generator, which generates a number $0\leq r<1$.  If $r\leq
\kappa_{ij} dt$ then a transition occurs.  Such an algorithm
corresponds to a two state Markov process and gives the proper
statistics so long as $dt$ is chosen such that $\kappa_{ij} dt \ll 1$,
for $i \neq j$.

Given the above algorithm, the equation of motion is integrated and
the position of the probe is determined.  Initially case~A is only
considered, where the lifetime of survival is much greater than the
intrinsic time constant of the experimental probe.  Here, traditional
methods (like those discussed here) are applicable.

\begin{figure}
\begin{center}
\centerline{\epsfxsize=3.2truein \epsfbox{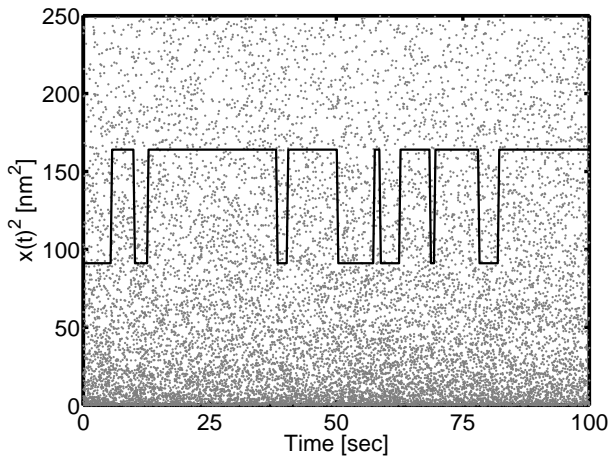}} 
\caption{The figure displays the instantaneous squared displacement
($x(t)^2$) of the experimental probe, gray dots.  The rates of
transition between the states are $\kappa_{12}$ = 0.26~Hz and
$\kappa_{21}$ = 0.11~Hz.  The solid bold line displays the mean
squared displacement ($\langle x^2\rangle = \kT/(k_{\rm P}+k_{{\rm
m}i})$) of the probe for the given state.}\label{fig:instant}
\end{center}
\end{figure}

Because instantaneous measurements cannot decipher the state of the
system, identification of the states have relied on averaging
procedures.  To illustrate the need for some type of averaging
procedure the instantaneous squared displacement of the probe $x(t)^2$
is displayed in \fref{fig:instant} (gray dots).  Instantaneous
measurements do not provide enough information to reveal the state of
the system and are not accurate predictors for the mean squared
displacement of the probe.  The mean squared displacement for state
$i$ corresponds to $\langle x^2\rangle_i = \kT/(k_{\rm P}+k_{{\rm
m}i})$ and is displayed as a black solid line in the figure.

\begin{figure}
\begin{center}
\centerline{\epsfxsize=3.2truein \epsfbox{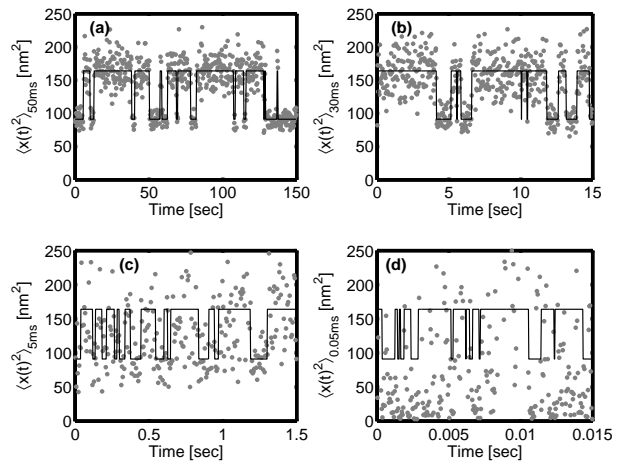}} 
\caption{The squared displacements averaged, $\langle
x(t)^2\rangle_{\rm T}$ (gray dots), over a small time interval T.
The solid bold lines correspond to the mean squared displacement
($\langle x^2\rangle = \kT/(k_{\rm P} + k_{{\rm m}i})$) of the probe
for the given state.  (a) Case A: $\kappa_{12}$ = 0.26~Hz,
$\kappa_{21}$ = 0.11~Hz and T = 100 milliseconds. (b) Case B:
$\kappa_{12}$ = 2.6~Hz, $\kappa_{21}$ = 1.1~Hz and T = 50
milliseconds. (c) Case C: $\kappa_{12}$ = 26~Hz, $\kappa_{21}$ = 11~Hz
and T = 5 milliseconds.  (d) Case D: $\kappa_{12}$ = 2600~Hz,
$\kappa_{21}$ = 1100~Hz and T = 0.05 milliseconds.}\label{fig:average}
\end{center}
\end{figure}

As argued, when averaged over a number of uncorrelated measurements M
(M $\approx \text{T}/\tau_{\rm P}$) can accurate predictions for the
mean squared displacement (second moment) be obtained, revealing
properties of the biomolecular state.  In addition, if this time
interval small compared to the lifetimes of the states ($\tau_{\rm P}
\ll \text{T} \ll \{\mathcal{T}_i\}$) then the individual states of the
system can be identified.  To exemplify this point \fref{fig:average}
(a) displays the squared displacement of the probe $\langle
x(t)^2\rangle_{\rm T}$, for case A, where the squared displacements
are averaged over a 50 millisecond time interval (number of
uncorrelated measurements M $\approx 100$).  Here the gray dots
correspond to the time averaged value $\langle x(t)^2\rangle_{\rm T}$
while the black solid line corresponds to the mean squared
displacement $\langle x^2\rangle_i$ for that particular state.  Unlike
the results in \fref{fig:instant}, now the state of the molecule can
be identified at time t and the properties of the states can be
determined.  The mechanical properties can be determined by
calculating the squared displacement in each state and setting it to
$\kT/(k_{\rm P} + k_{{\rm m}i})$ and the lifetimes of the states can
be determined by accumulating the survival time of each state.

We now explore cases where the rates are increased by orders of
magnitude in order to determine the limitations of the above averaging
procedure.  Initially, the rates are increased by an order of
magnitude, corresponding to lifetimes of 380~ms and 910~ms (case B).
Figure~\ref{fig:average}~(b) displays the averaged squared
displacement $\langle x(t)^2\rangle_{\rm T}$ where the time window for
averaging T is taken to be 30~ms (M $\approx 60$).  Here too, we are
in the regime where the lifetimes of the states are much greater than
the intrinsic time constant of the experimental probe and; therefore,
a time window can be taken such that the states can be accurately
identified and the properties of the states can be determined.
Results for increasing the rate by another order of magnitude,
corresponding to lifetimes of 38~ms and 91~ms (case C), are displayed
in \fref{fig:average} (c).  The squared displacement is averaged over
a time window T = 5~ms (M $\approx 10$) which is an order of magnitude
larger than the intrinsic time constant of the experimental probe.
For this case it becomes extremely difficult to identify the states of
the system.  Finally, case D is displayed in \fref{fig:average} (d),
where the rates are now increased by an additional two orders of
magnitude.  Here the rates are of order of the inverse intrinsic time
constant of the experimental probe.  The squared displacements are
averaged over a time window T = 0.05~ms (M $\approx 1$).  Here too,
the states of the system cannot be resolved when using this
conventional technique.

The failure to identify the states and their properties in case C
(\fref{fig:average}(c)) and case D (\fref{fig:average}(d)) lies in the
fact that the time window of averaging T is not large enough to
contain a significant number of uncorrelated measurements.  It may be
envisioned that this failure could be alleviated if the time window of
averaging T was increased.  However, to accurately identify the states
of the system (and hence determine their mechanical and kinetic
properties) the likelihood of a transition occurring within this
measuring period T must be small.  This requires a time window an
order of magnitude less than (and generally much less than) the
lifetime of the states, T~$ \ll \{\mathcal{T}_i\}$.  Increasing the
time window will result in a number of measurements corresponding to
mixed states, not providing better statistics regarding the properties
of the individual states.

Another possible route in attempting to alleviate the failure, in
identifying the properties of the states for case C and D, might rely
on increasing the frame rate of observation.  By frame rate of
observation we mean the rate at which the instantaneous measurements
are stored.  Upon increasing results in a larger set of data points
to calculate the second moment $\langle x(t)^2\rangle_{\rm T}$ over
fixed time window T. \ However, increasing the frame rate of
observation does not reduce the noise, because it does not increase
the number of uncorrelated measurements.  Many uncorrelated
measurements are needed to obtain good statistics and it takes time of
order $\tau_{\rm P}$ to make two uncorrelated measurements.  When
averaged over the same time window T the features of the plots in
\fref{fig:average} are not altered when the frame rate of
observation is increased.

We have illustrated that to resolve the dynamics and statistics of a
biomolecule, which is undergoing structural transitions, requires an
experimental apparatus whose intrinsic time constant is much smaller
than the lifetimes of the states, $\tau_{\rm P} \ll
\{\mathcal{T}_i\}$.  This requirement is based upon the precondition
that the properties of the individual states are to be identified
separately; requiring an experimental probe to have sufficient
temporal resolution to adequately explore each state prior to the
occurrence of a transition.  One of the key ambitions of the remainder
of the paper is to develop a formalism that allows us to forgo this
precondition and in turn alleviate the demand that $\tau_{\rm P} \ll
\{\mathcal{T}_i\}$.

To alleviate this demand we propose an alternative analytic approach
for analyzing the data from single-molecule probe techniques.  The
property that we compute is the position-position auto-correlation
function.  To illustrate the potential benefit of analyzing the
auto-correlation function, we now present numerical solutions for
sample cases presented in this section.  This analysis will lend
insight into how the form of the auto-correlation function is
influenced by the rates of transition between biomolecular states and
how such properties are revealed.  Following this analysis we present
our general mathematical theory.

\subsection{Numerical solution for the auto-correlation function}\label{sec:numac}

We present numerical solutions for the auto-correlation function for
two of the cases just analyzed, case~B~and~D.  The auto-correlation
function is defined as
\begin{eqnarray}
\av{\delta x(t)\delta x(0)} &=& \av{x(t)x(0)}  - \av{x}^2, \label{eqn:ac}
\end{eqnarray}
where $\delta x(t)$ is the displacement of the probe from the
equilibrium value ($\langle x\rangle$) at time $t$, 
$$
\delta x(t) = x(t) - \langle x\rangle.
$$  The statistical averages in
\eref{eqn:ac} can be interpreted in terms of time averages.  The
first term on the right-hand side is defined as
\begin{eqnarray}
\langle x(t)x(0)\rangle &=& \lim_{\text{T} \rightarrow \infty}
 \frac{1}{\text{T}}\int_0^\text{T} dt' \ x(t+t')x(t')
\end{eqnarray}
and the equilibrium position of the probe is defined as
\begin{eqnarray}
\av{x} &=& \lim_{\text{T} \rightarrow \infty}\frac{1}{\text{T}}\int_0^\text{T} dt' \ x(t').
\end{eqnarray}

The auto-correlation function measures the memory of particle
displacements.  Given an initial displacement at time $t = 0$, the
auto-correlation function gives the time scale at which the particle
returns to its equilibrium value and looses accounts of that given
initial displacement.  If the likelihood of a transition between
biomolecular states is high during this relaxation period, then upon
return to its equilibrium position the particle itself should have
memory regarding the various states visited during its journey;
therefore, such transitions should reveal themselves through the
solution to the auto-correlation function.  The auto-correlation
function decays on time scales of the order of the intrinsic time
constant of the experimental probe, potentially capable of revealing
transitions occurring on these time scales.

\begin{figure}
\begin{center}
\centerline{\epsfxsize=3.2truein \epsfbox{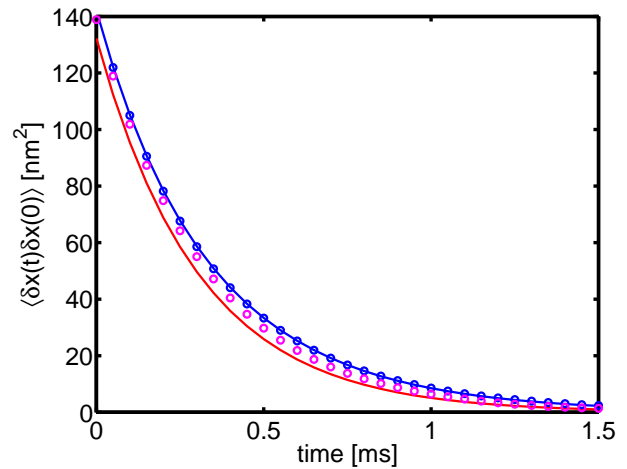}} 
\caption{Auto-correlation function for a system with a fluctuating
spring constant.  Case~B and case~D are represented by blue circles
and red circles, respectively.  The solid lines refer to physically
motivated guesses for the solution of the auto-correlation function.
These guesses are expressed in \erefs{eqn:ssg}~and~\ref{eqn:fsg}, blue
line and red line, respectively.}\label{fig:ac_num}
\end{center}
\end{figure}

To explore this possibility, we now present numerical solutions for
the auto-correlation function for case~B and case~D.  Given the
numerical solution for the position of the probe ($x(t)$) the
auto-correlation function can be determined through application of
\eref{eqn:ac}.  The results are displayed in
\fref{fig:ac_num}, where the solution for case~B is displayed as
blue circles and that for case~D is displayed as red circles.
(Recall, case~B corresponds to a system where transitions occur on
time scales that are large compared to the intrinsic time constant of
the experimental probe ($\tau_{\rm P} \ll \{\mathcal{T}_i\}$), while
case~D corresponds to a system where transitions occur on the time
scale on the order of the intrinsic time constant of the experimental
probe ($\tau_{\rm P} \approx \{\mathcal{T}_i\}$).)  The solution for
both cases differ, which is solely due to difference in the magnitude
of the rates of transition, see Table~\ref{tab:rates}.

To garner insight into how the solutions for the auto-correlation
functions depends on the magnitude of the rates of transition between
biomolecular states, we now present two physically motivated guesses
for the solution.  We term these two guesses as the slow-switching
limit and fast-switching limit solutions.  (These solutions will be
derived mathematically in \sref{sec:limits}.)  If the molecule
switches on time scales that is ``long'' compared to the intrinsic
time constant of the experimental probe, then upon returning to is
equilibrium position the probe will only observe a single biomolecular
state.  Upon averaging over many trajectories, switching will be
infrequent and the probe should be capable of exploring each states
statistical properties.  The auto-correlation function should then
just be the average auto-correlation function of each state
\begin{equation}
\langle\delta x(t)\delta x(0)\rangle =
\Big\langle\frac{\kT}{k}e^{-t/\tau}\Big\rangle,
\label{eqn:ssg}
\end{equation}
where explicitly
$$ \Big\langle\frac{\kT}{k}e^{-t/\tau}\Big\rangle = \sum_i P_i
\frac{\kT}{k_i}e^{-t/\tau_i}.
$$ Here $P_i$ is the probability that the molecule is in state $i$,
$k_i$ is the associate spring constant of state $i$ (\eref{eqn:ki})
and $\tau_i$ is the time constant of the probe when the biomolecule is
in state $i$ ($\tau_i = \gamma/k_i$).  This solution is termed the
slow-switching limit solution, because the molecules switching
(between states) is slow compared to the motion of the probe.  On the
other hand, if the molecule switches between states on time scales
that are very ``short'' compared to the intrinsic time constant of the
experimental probe, then the molecule equilibrates between all states
on time scales much shorter than those governing the motion of the
probe.  Therefore, the probe ``sees'' the average response of the
molecule.  The solution for the auto-correlation function should then
correspond to that of an equilibrated system with the spring constant
equal to the average spring constant of the biomolecule,
\begin{equation}
\langle\delta x(t)\delta x(0)\rangle = \frac{\kT}{\langle
k\rangle}e^{-t\langle \tau^{-1}\rangle}\label{eqn:fsg}
\end{equation}
where
\begin{equation}
\langle \tau^{-1}\rangle = \langle k\rangle/\gamma\label{eqn:taukave}
\end{equation}
is the mean $k$-weighted~\footnote{We make an explicit distinction
between the average time constant $\av{\tau} = \sum_iP_i\tau_i$ and
the inverse of the mean $k$-weighted inverse time constant
$1/\avtauinv = \gamma/\av{k}$.}  inverse time constant.  This solution
is termed as the fast-switching limit solution, as the molecules
switching (between states) is fast compared to the motion of the
probe.  In \sref{sec:limits} we show that these solutions are indeed
solutions for the auto-correlation function in the limit that the
molecule switches either ``slow'' or ``fast'' compared to the motion
of the probe.

Figure~\ref{fig:ac_num} displays these representations for the
auto-correlation function.  The slow-switching limit solution,
\eref{eqn:ssg}, is displayed as a blue line and the fast-switching
limit solution, \eref{eqn:fsg}, is displayed as a red line.  The
auto-correlation function for case~B lays on the physically motivated
slow-switching limit solution, \eref{eqn:ssg}.  This is expected,
because case~B corresponds to a system where the switching between the
biomolecule states is ``slow'' compared to the decay time of the
experimental probe.  The time scales governing the switching for
case~D lies somewhere in between the mentioned extremes.  Therefore,
it is reasonable to expect its solution to lie in between these two
limits, as depicted in the figure.

Although the solutions for the auto-correlation function for both
case~B and case~D differ, their separation is only a few percent of
their solution.  From an experimental point of view it is important to
understand the magnitude and origin of this separation, so as to
determine whether or not an experiment can provide sufficient spatial
resolution to differentiate between various cases.  To gain insight
into how this separation may be influenced by the mechanical response
of the biomolecule, we now expand our parameter space and consider two
different cases, termed case~$\mathcal{B}$ and case~$\mathcal{D}$.
The parameters now governing the biomolecule (rates and spring
constants) are the same as their counterparts (case~B~and~D) except
that we choose the spring constant for the biomolecule in state 1 to
be larger, $k_{{\rm m}1}$ = 120 pN/$\mu$m.  (Recall, for case~B~and~D
$k_{{\rm m}1}$=25~pN/$\mu$m, $k_{{\rm m}2}$=5~pN/$\mu$m and $k_{\rm
P}$=20~pN/$\mu$m.) Although fairly larger, this new spring constant is
still easily obtainable in muscle
proteins~\cite{rief97b,oberhauser98,marszalek99,rief99,oberhauser01}.

\begin{figure}
\begin{center}
\centerline{\epsfxsize=3.2truein \epsfbox{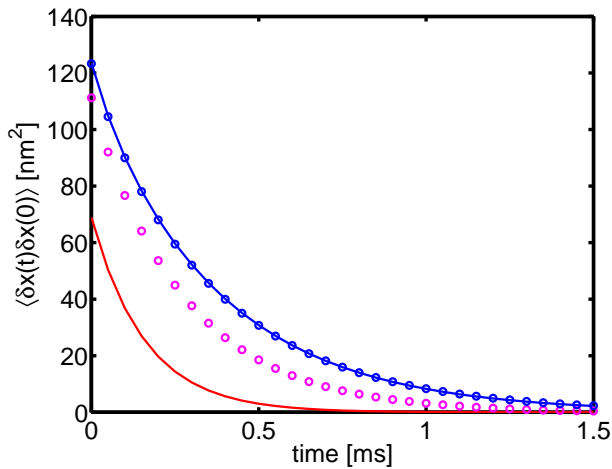}} 
\caption{Auto-correlation function for a system with a fluctuating
spring constant.  The parameters used to define the system here are
similar to there counterparts displayed in \fref{fig:ac_num},
except the spring constant for the biomolecule in state 1 is
120~pN/$\mu$m.  Case~$\mathcal{B}$~and~$\mathcal{D}$ are represented
by blue circles and red circles, respectively. The solid lines refer
to physically motivated guesses, expressed in
\erefs{eqn:ssg}~and~\ref{eqn:fsg} (blue line and red line,
respectively).}\label{fig:ac_num_largek1}
\end{center}
\end{figure}

Figure~\ref{fig:ac_num_largek1} displays the results for the
auto-correlation function in this new example.  The separation between
the two cases ($\mathcal{B}$~and~$\mathcal{D}$) is now much larger
than the results presented (B~and~D) in \fref{fig:ac_num}.
Additionally, there is a larger separation between the two physically
motivated solution, represented by \erefs{eqn:ssg}~and~\ref{eqn:fsg}.
The increase is a direct consequence of the increase in the difference
between the spring constants controlling the motion of the probe
(difference between $k_1$ and $k_2$, $k_i = k_{{\rm m}i} + k_{\rm
P}$).  General features regarding the influence of the biomolecular
properties on spatial separation is further explored in
\sref{sec:convergence}.

Although there is an increase in the separation between various
solutions the cousin of case~B, case~$\mathcal{B}$, also lies on the
form expressed in \eref{eqn:ssg}; supporting the notion that this form
is the proper solution for the auto-correlation function when the
switching between biomolecular states is ``slow'' compared to the time
scales governing the motion of the experimental probe.  As the rates
increase (going from case~$\mathcal{B} \rightarrow$
case~$\mathcal{D}$) the solution for the auto-correlation function
approaches the solution represented by \eref{eqn:fsg}, similar to
results found in \fref{fig:ac_num}.  This is consistent with the
interpretation that the form of the auto-correlation function
represented by \eref{eqn:fsg} corresponds to the case where the
biomolecule switches between states on times scales that are much
faster than those governing the motion of the experimental probe.

\begin{figure}
\begin{center}
\centerline{\epsfxsize=3.2truein \epsfbox{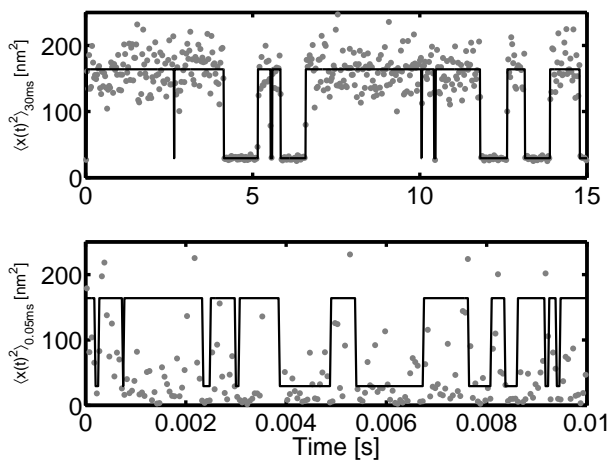}} 
\caption{Mean squared displacement of the probe ($\langle
x(t)^2\rangle_{\rm T}$) averaged over some time window T, represented
by gray dots.  The top figure is the analysis for case~$\mathcal{B}$
(T=30~ms) and bottom figure is for case~$\mathcal{D}$ (T=0.05~ms).
Solid black line represents the mean squared displacement of the probe
in that particular state.}\label{fig:average_Largek1}
\end{center}
\end{figure}

Although for these new cases the spatial resolution (difference
between the two solutions) for the auto-correlation function has
increased, limitations imposed by traditional methods (regarding time
resolution) have not altered.  Figure~\ref{fig:average_Largek1}
displays the traditional averaging procedure presented earlier in this
section.  Again, the properties of the biomolecule when represented by
case~$\mathcal{B}$ can be adequately identified with this method,
those represented by case~$\mathcal{D}$ cannot.
 
In this section we presented numerical solutions for the position of
an experimental probe and its associated auto-correlation function,
when the probe is attached to a biomolecule with fluctuating
mechanical response.  It was demonstrated that when the biomolecule
fluctuates between states on time scales governing the motion of the
experimental probe, sufficient temporal resolution is not provided for
traditional analysis to be amiable for identifying the mechanical and
kinetic properties of the biomolecule.  Furthermore, we demonstrated
that in this temporal regime the solution for the auto-correlation
function is {\em affected} by the biomolecular properties (rates and
mechanical response) and can potentially be used in the analysis of
experimental data.

Our numerical study provided insight into the problem of interest.
Although, they proved to be valuable in this context, numerical
solutions for the equation of motion are not always the most efficient
theoretical means to analysis experimental data; particularly when
biomolecular parameters of interest (rates and spring constants) are
to be determined from the data itself (by least squares fit for
example).  For such cases analytic solutions provide a more promising
alternative.  We now proceed with the mathematical formalism for
calculating the auto-correlation function for a system fluctuating
between mechanically distinct states.  Initially, we present bounds
for the general solution to the auto-correlation function.  Although,
no general ``closed'' form solution for the auto-correlation function
is known (at least to our knowledge) we present solutions in the form
of two different series expansions, that are expressed around the two
opposing bounds.  Later, we present a general analysis regarding when
the form of the auto-correlation function is appreciably influenced by
the biomolecular properties.

\section{Theoretical Formalism:  Motion of an over-damped particle in a switching harmonic trap; bounds on the auto-correlation function} \label{sec:theory}

We now present our theoretical framework for calculating the
auto-correlation function for a system that is in a thermal heat bath
and is undergoing transitions between various mechanical states.  In
this section we prove that the two physically motivated solutions
presented in the previous section are indeed bounds to the general
solution of the auto-correlation function.  These bounds will
determine the spatial resolution of the problem at hand, by providing
the possible range over which the auto-correlation function can
fluctuate.  In addition, these bounds serves are starting points for
series solutions for the auto-correlation function presented in
\sref{sec:expansion}.

To make this analysis more manageable (though not changing the
principle of the calculation) we now assume that the equilibrium
position is the same for all of the states and hence $x^o(t) = 0$.  In
single molecule experiments this would correspond to measuring the
motion of the experimental probe perpendicular to the orientation of
the biomolecule, for example.  The equation of motion,
\eref{eqn:case1}, now becomes
\begin{equation}
\gamma \dot x(t) =  F_r(t) - k(t) x(t). \label{eqn:caseN}
\end{equation}
Additionally, we impose the stationary condition for the fluctuating
spring constant.  The stationary condition states that the average
value of an arbitrary function of the fluctuating spring constant is
independent of time translations.  For a function $G(k(t))$ that is
dependent on the spring constant at a single instant of time, the
stationary conditions states that its average is independent of time,
\begin{equation}
\av{G(k(t))} =
\av{G(k)}. \label{eqn:statcond}
\end{equation}

In addition to the stationary condition we remind the reader that the
assumptions presented in \sref{sec:background} still hold.  These
assumptions lead to the following important fact.  Because the random
thermal force is Gaussian and its first two moments
(\erefs{eqn:fm}~and~\ref{eqn:sm}) are independent of the state of the
molecule, the random thermal force ($F_r(t)$) is independent of the
state of the molecule and; therefore, statistically independent of the
spring constant $k(t)$.  (This assumption does not hold if the damping
constant was state dependent.)  This results in a decoupling when
averaging over functions of the random force and spring constant,
\begin{eqnarray}
\av{G(k(t),\cdots,k(t'))H(F_r(s),\cdots,F_r(s'))} = &&\label{eqn:decouple}\\
\av{G(k(t),\cdots,k(t'))}\av{H(F_r(s),\cdots,F_r(s'))},&&\nonumber
\end{eqnarray}
where $G$ and $H$ are arbitrary functions.  This fact will become
important for the analysis of the auto-correlation function.

\subsection{Formal Solution for position of the probe}

To obtain bounds for the general solution to the auto-correlation
function we first present a formal representation for the solution to
the position of the experimental probe.  Equation~\ref{eqn:caseN} is
a linear, inhomogeneous, stochastic, differential equation with
stochastic
coefficients~\cite{bourret73,brissaud74,vankampen75,roerdink81,lindenberg81,roerdink82,ikeda00}.
The general solution to \eref{eqn:caseN} (i.e. the position of
the experimental probe) can be written formally as
\begin{equation}
x(t) = \frac{1}{\gamma}\int_0^\infty
e^{-\int_0^{t'}k(t-t'')/\gamma \ dt''}F_r(t-t')dt', \label{eqn:xtg}
\end{equation}
where $k(t)$ is the fluctuating spring constant.  What lays difficulty
into this problem is that the exponential is stochastic, due to the
fluctuating spring constant.  Evaluation of such a term and the
associated averages are quite difficult, as we are aware of no general
solution.  Only special cases have been
solved~\cite{bourret73,brissaud74,lindenberg81,ikeda00}.  Although no
general solution to this problem exists, the formal solution for the
position of the probe, \eref{eqn:xtg}, can be used to obtain
bounds on the auto-correlation function.

\subsection{Bounds on the auto-correlation function}\label{sec:bounds}

Using the formal solution for the position of the probe we now provide
an upper and lower bound to the general solution for the
auto-correlation function.  These bounds correspond to the two
physically motivated solutions we presented in the previous section,
\erefs{eqn:ssg}~and~\ref{eqn:fsg}.  Given the formal solution for the
position of the probe, \eref{eqn:xtg}, the solution for the
auto-correlation function is expressed as
\begin{widetext}
\begin{eqnarray}
\av{\delta x(t)\delta x(0)} &=&  \Big\langle
\frac{1}{\gamma^2}\int_0^\infty\int_0^\infty  dt' ds'
e^{-\int_0^{t'} dt''  k(t-t'')/\gamma}e^{-\int_0^{s'} ds'' 
k(-s'')/\gamma} F_r(t-t')F_r(-s')\Big\rangle  \nonumber \\&=&  
\frac{1}{\gamma^2}\int_0^\infty\int_0^\infty  dt'  ds' 
\Big\langle e^{-\int_0^{t'} dt''  k(t-t'')/\gamma}e^{-\int_0^{s'} ds'' 
k(-s'')/\gamma}F_r(t-t')F_r(-s')\Big\rangle.
\end{eqnarray}
\end{widetext}
We now exploited the fact that within our assumptions that the random
force is statistically independent of the fluctuating spring constant,
\eref{eqn:decouple}.  The general solution to the
auto-correlation function is now expressed as
\begin{widetext}
\begin{eqnarray}
\av{\delta x(t)\delta x(0)} &=&
\frac{1}{\gamma^2}\int_0^\infty\int_0^\infty dt' ds' \Big\langle
e^{-\int_0^{t'} dt'' k(t-t'')/\gamma}e^{-\int_0^{s'} ds''
k(-s'')/\gamma}\Big\rangle\Big\langle F_r(t-t')F_r(-s')\Big\rangle \\
&=& \frac{1}{\gamma^2}\int_0^\infty\int_0^\infty dt' ds' \Big\langle
e^{-\int_0^{t'} dt'' k(t-t'')/\gamma}e^{-\int_0^{s'} ds''
k(-s'')/\gamma}\Big\rangle2\kT\gamma\delta(s'-(t'-t)),
\end{eqnarray}
\end{widetext}
where the definition of the second moment for the random force,
\eref{eqn:sm}, has been used.  Carrying out the integration over
the dummy variable $s'$ the auto-correlation function is
\begin{eqnarray}
&&\av{\delta x(t)\delta x(0)} = \label{eqn:acgen}  \\&&  
\frac{2\kT}{\gamma}\int_t^\infty  dt'   
\Big\langle e^{-\int_0^{t'} dt''  k(t-t'')/\gamma}e^{-\int_0^{t'-t} ds'' 
k(-s'')/\gamma}\Big\rangle.\nonumber
\end{eqnarray}

We are now in position to obtain bounds for the auto-correlation
function.  To obtain an upper bound we start with the general
definition of the auto-correlation function, as expressed in
\eref{eqn:acgen}, and shift the last integral in the exponent by $t'$
and change the dummy variable to $t''$.  The general solution for the
auto-correlation function is now expressed as
\begin{eqnarray}
&&\av{\delta x(t)\delta x(0)} = \\&& \frac{2\kT}{\gamma}\int_t^\infty
dt' \Big\langle e^{-\int_0^{t'} dt'' k(t-t'')/\gamma}e^{-\int_{t'}^{2t'-t}
dt'' k(t'-t'')/\gamma}\Big\rangle.\nonumber
\end{eqnarray}

To simplify our expression we define a function $g(t,t',t'')$, which
has the property
\begin{equation}
\gamma g(t,t',t'') = \left\{\begin{array}{ll} k(t-t'') & \ \ t'' \leq t'
\\k(t'-t'') & \ \ t'' > t' \end{array}\right.. \label{eqn:gsttp}
\end{equation}
The auto-correlation function may now be expressed with only one
integral in the exponent,
\begin{equation}
\av{\delta x(t)\delta x(0)} = \frac{2\kT}{\gamma}\int_t^\infty dt'
\Big\langle e^{-\int_{0}^{2t'-t} dt''  g(t,t',t'')}\Big\rangle.\label{eqn:aculd}
\end{equation}
For a given realization of the molecule over time (a particular
contribution to the average) an upper bound can be placed on the
exponential
\begin{equation}
e^{-\int_{0}^{2t'-t} dt'' g(t,t',t'')} \leq
\frac{1}{2t'-t}\int_{0}^{2t'-t}dt'' \ e^{-(2t'-t)g(t,t',t'')}.
\end{equation}
This bound results because the arithmetic mean is greater than the
geometric mean~\cite{intserprod}.  This inequality leads to an upper
bound on the auto-correlation function,
\begin{eqnarray}
&&\av{\delta x(t)\delta x(0)} \leq \label{eqn:aculd_b} \\ &&
\frac{2\kT}{\gamma}\int_t^\infty dt' \frac{1}{2t'-t} \int_{0}^{2t'-t}
dt'' \Big\langle e^{-(2t'-t)g(t,t',t'')}\Big\rangle,\nonumber 
\end{eqnarray}
where we have taking the average inside of the inner integral.  To
obtain our desired form for the upper bound we exploit the definition
of $g(t,t',t'')$ (\eref{eqn:gsttp}) and the stationary condition
(\eref{eqn:statcond}).  From the definition of $g(t,t',t'')$ the
inner integral is expressed as
\begin{widetext}
\begin{equation}
\int_0^{2t'-t} dt'' \Big\langle e^{-(2t'-t)g(t,t',t'')}\Big\rangle =
\int_0^{t'} dt'' \Big\langle e^{-(2t'-t)k(t-t'')/\gamma}\Big\rangle +
\int_{t'}^{2t'-t} dt'' \Big\langle
e^{-(2t'-t)k(t'-t'')/\gamma}\Big\rangle.
\end{equation}
\end{widetext}
The stationary condition implies that the terms within the averages
are independent of time, resulting in the following simplification
\begin{widetext}
\begin{eqnarray}
\int_0^{t'} dt'' \Big\langle e^{-(2t'-t)k(t-t'')/\gamma}\Big\rangle +
\int_{t'}^{2t'-t} dt'' \Big\langle
e^{-(2t'-t)k(t'-t'')/\gamma}\Big\rangle &= & \int_0^{t'} dt''
\Big\langle e^{-(2t'-t)k/\gamma}\Big\rangle + \int_{t'}^{2t'-t} dt''
\Big\langle e^{-(2t'-t)k/\gamma}\Big\rangle \nonumber \\ &=& t'
\Big\langle e^{-(2t'-t)k/\gamma}\Big\rangle + (t'-t)\Big\langle
e^{-(2t'-t)k/\gamma}\Big\rangle \nonumber \\ &=& (2t'-t) \Big\langle
e^{-(2t'-t)k/\gamma}\Big\rangle. \label{eqn:intfun}
\end{eqnarray}
\end{widetext}
Substituting this result into \eref{eqn:aculd_b}, the bound on
the auto-correlation function is now
\begin{eqnarray}
\av{\delta x(t)\delta x(0)} \leq \frac{2\kT}{\gamma}\int_t^\infty dt'
\Big\langle e^{-(2t'-t)k/\gamma}\Big\rangle.
\end{eqnarray}
Performing the integral we have
\begin{eqnarray}
\av{\delta x(t)\delta x(0)} &\leq& \Big\langle
\frac{\kT}{k}e^{-t/\tau}\Big\rangle.
\end{eqnarray}
The right hand term is indeed the slow-switching limit solution for
the auto-correlation function.  Proving that the slow-switching limit
solution is an upper bound for the auto-correlation function.

Having obtained an upper bound for the auto-correlation function we
now proceed to obtain a lower bound.  We now return to the general
definition of the auto-correlation function, \eref{eqn:acgen},
and express the fluctuating spring constant in terms of its average
value and the fluctuations around that average,
\begin{equation}
k(t) = \av{k} + \delta k(t) \label{eqn:ktd}
\end{equation}
where
$$ 
\delta k(t) \equiv k(t) - \av{k}.
$$ Due to the stationary condition the average value of the spring
constant is time independent.  With this new expression for the
fluctuating spring constant, the auto-correlation function is now
expressed as
\begin{widetext}
\begin{eqnarray}
\av{\delta x(t)\delta x(0)} &=& \frac{2\kT}{\gamma}\int_t^\infty
dt' e^{\langle k\rangle t/\gamma}e^{-2\langle k\rangle t'/\gamma}
\Big\langle e^{-(\int_0^{t'} dt'' \delta k(t-t'')/\gamma + \int_0^{t'-t}
ds'' \delta k(-s'')/\gamma)}\Big\rangle.\label{eqn:xxbound_a}
\end{eqnarray}
\end{widetext}

The average in \eref{eqn:xxbound_a} is a sum over all possible
realizations of the molecule.  For a given realization, the argument
of the exponential is a fixed number, which allows us to exploit the
inequality
\begin{equation}
e^X \geq 1 + X, \label{eqn:exp}
\end{equation}
for arbitrary $X$.  With this inequality a lower bound can now be
placed on the auto-correlation function
\begin{widetext}
\begin{eqnarray}
\av{\delta x(t)\delta x(0)} &\geq& \frac{2\kT}{\gamma}\int_t^\infty
dt' e^{\langle k\rangle t/\gamma}e^{-2\langle k\rangle t'/\gamma}
\Big\langle 1 - \Big(\int_0^{t'} dt'' \delta k(t-t'')/\gamma +
\int_0^{t'-t} ds'' \delta k(-s'')/\gamma\Big)\Big\rangle\nonumber \\
&\geq& \frac{2\kT}{\gamma}\int_t^\infty dt' e^{\langle k\rangle
t/\gamma}e^{-2\langle k\rangle t'/\gamma} \left(1 - \Big(\int_0^{t'} dt''
\Big\langle\delta k(t-t'')\Big\rangle/\gamma + \int_0^{t'-t} ds''
\Big\langle\delta k(-s'')\Big\rangle/\gamma\Big)\right).\label{eqn:xxbound_b}
\label{eqn:xxbound_b}
\end{eqnarray}
\end{widetext}
Exploiting the stationary condition, \eref{eqn:statcond}, the
later terms vanish
\begin{eqnarray*}
\av{\delta k(t)} &=& \av{k(t)} - \av{k} \\
&=& \av{k}-\av{k}\\
&=& 0.
\end{eqnarray*}
Therefore, the lower bound expressed in \eref{eqn:xxbound_b} can
be written as
\begin{eqnarray}
\av{\delta x(t)\delta x(0)} &\geq& \frac{2\kT}{\gamma}\int_t^\infty
dt' e^{\langle k\rangle t/\gamma}e^{-2\langle k\rangle t'/\gamma}.
\end{eqnarray}
Performing the integrals we get our desired result
\begin{eqnarray}
\av{\delta x(t)\delta x(0)} &\geq& \frac{\kT}{\langle
k\rangle}e^{-t\langle \tau^{-1}\rangle}.
\end{eqnarray}
The lower bound is the fast-switching limit solution and has the
physical interpretation that due to the fluctuations in the spring
constant over time, the displacement of the probe is more correlated
to past events than an equivalent system with a single spring constant
equal to the average spring constant of the fluctuating system.  The
auto-correlation function not only reflects the average response of
the biomolecule but also correlations between the biomolecular states.

The fact that these two solutions bound the general solution to the
auto-correlation function,
\begin{equation}
\frac{\kT}{\langle k\rangle}e^{-t\langle \tau^{-1}\rangle} \leq
\av{\delta x(t)\delta x(0)} \leq \Big\langle
\frac{\kT}{k}e^{-t/\tau}\Big\rangle, \label{eqn:bounds}
\end{equation}
also appeals to physical intuition.  Consider the fanciful yet
instructive case where the magnitude of the rates for the biomolecule
can be tuned, similar to the analysis presented in
\sref{sec:background}.  Consider a parameter $\alpha$ which describes
this tuning, $\{\kappa_{ij}\} \rightarrow \{\alpha\kappa_{ij}\}$, for
$i \neq j$.  If we initially set $\alpha \ll 1$ then the solution for
the auto-correlation function will correspond to the slow-switching
limit.  As the parameter $\alpha$ increases, switching occurs more
quickly and the solution for the auto-correlation function deviates
from the slow-switching limit solution and begins to approach the
solution in the fast-switching limit. As the parameter further
increases, such that $\alpha \gg 1$, we are in the fast-switching
limit.  (This type of tuning is displayed in
\frefs{fig:ac_num}~and~\ref{fig:ac_num_largek1}.)  As we tune this
parameter $\alpha$ the solution for the auto-correlation function
starts at one limit and approaches the other, always lying in between
these two bounds.

In this section we have obtained an upper and lower bound for the
general solution to the auto-correlation function.  These bounds were
found by exploiting properties of the formal solution,
\eref{eqn:acgen}.  However, the value of the formal solution as a tool
amiable to the analysis of experimental data is less evident.
Transforming the formal solution into a tractable ``closed'' form
proves to be difficult, as we are aware of no such solutions.
Solutions more tractable for the analysis of experimental data can be
sought by a series expansion~\cite{vankampen75,roerdink81,roerdink82}.
In this sense, we follow earlier developments, though our method
differs from previous methods in that the series is expanded around
two important experimental limits: the fast- and slow-switching
limits.

A great benefit of presenting these two series solutions is that they
approach the exact solution from opposing bounds.  To generate these
two series solutions we choose to first present a single series
solution for the auto-correlation function, upon which both limiting
forms can be derived.  Upon deriving the limiting forms we transform
the single series into the two desired series solutions.  Generating
the two series solutions in this fashion will lead to a robust
analysis of the error associated with truncating the series at a given
order (\sref{sec:convergence}).  Giving a prediction regarding the
convergence of each series {\em a priori}.

\section{Series Solution for the Auto-Correlation Function}~\label{sec:expansion}

We now present a series solution for auto-correlation function, upon
which the two series solutions which initiate at the limiting forms
will be derived.  To determine this first series solution we express
the solution for the position of the probe $x(t)$ itself in the form
of a series.  Given the solution for the position of the probe $x(t)$,
the auto-correlation function can be determined.  The position of the
probe $x(t)$ satisfies the equation of motion represented in
\eref{eqn:caseN} and obtaining its solution is a formidable task as
the equation of motion depends on two stochastic variables: $F_r(t)$
and $k(t)$.  To proceed we find it convenient to reformulate the
equation of motion into a slightly different form, whose utility will
be revealed below,
\begin{equation}
\gamma \dot x = F_r(t) - k_a x + f(t)k_d x. \label{eqn:BE}
\end{equation}
Now the switching between different mechanical states is captured in
the presence of the function $f(t)$.  By writing the dynamics in this
way, we are required to introduce new notation.  In particular, $k_a$
is the equally-weighted, average spring constant, averaged between the
maximum and minimum values,
\begin{equation}
k_a = (k_{\text{max}} + k_{\text{min}})/2,\nonumber
\end{equation}
$k_d$ is the equally-weighted, difference spring constant, which
reflects the difference between the maximum and minimum values,
\begin{equation}
k_d = (k_{\text{max}} - k_{\text{min}})/2,\nonumber
\end{equation}
and $f(t)$ is a random variable which determines the spring constant
of the system
\begin{equation}
f(t) = \frac{k_a - k_i}{k_d} \ \ \ \ \ \ \ \text{if in state $i$.}
\label{eqn:fi}
\end{equation}
We note that the choices for $k_a$ and $k_d$ are arbitrary though they
have the advantage that this choice bounds the magnitude of the
fluctuating variable $f(t)$ ($|f(t)| \leq$ 1) and the ratio of the
spring constants ($|f(t)k_d/k_a| \leq 1$) which provides analytical
advantages.

The new formulation can be interpreted as follows.  The probe is in a
harmonic trap with spring constant $k_a$, subjected to two random
forces: $F_r(t)$ mimics the thermal heat bath and $f(t)k_d$ emulates
the switching of the states of the molecule.  The form of the
switching term is now a small perturbation when compared to the
harmonic trap $k_a$, $|f(t)k_d/k_a| < 1$.  This switching term
``kicks'' the probe either toward the origin or away from it, depending
on whether state $i$ has a spring constant $k_i$ greater than or less
than $k_a$, respectively.

To obtain a series solution for the auto-correlation function we first
promote the fluctuating spring term, $-f(t)k_dx(t)$, in
\eref{eqn:BE} to the inhomogeneous part of the differential
equation (treat as an external force). Next standard solutions to the
problem can be used to formally rewrite the solution to
\eref{eqn:BE} as
\begin{equation}
x(t) = x_r(t) + x_f(t). 
\label{eqn:xt1}
\end{equation}
The first term $x_r(t)$ is the displacement induced by thermal
fluctuations in a harmonic well described by spring constant $k_a$,
namely,
\begin{equation}
x_r(t) = \frac{1}{\gamma}\int_0^\infty dt' e^{-\frac{t'}{\tau_a}}F_r(t-t'),
\label{eqn:xr}
\end{equation}
where $\tau_a = \gamma/k_a$ is the equally weighted average time
constant.  The latter term $x_f(t)$ is the displacement induced by the
fluctuating spring $f(t)k_d$ in the same harmonic well described by
spring constant $k_a$, namely,
\begin{equation}
x_f(t) = \frac{1}{\gamma}\int_0^\infty dt' e^{-\frac{t'}{\tau_a}}
f(t-t')k_dx(t-t'). \label{eqn:xf}
\end{equation}
The motion of the probe is now described by two types of random
``kicks''.  The thermal heat bath induces random ``kicks'' described
by motion $x_r(t)$ and the switching of the macromolecule induces
random ``kicks'' described by motion $x_f(t)$.

The solution for $x(t)$ is now a functional of itself (due to the
$x_f(t)$ term) and can be solved iteratively in terms of the
displacement due to thermal motion:
\begin{eqnarray}
&&x(t) = \label{eqn:xt} \\ && \sum_{n=0}^\infty
\left(\frac{1}{\tau_d}\right)^n\int_0^\infty 
\prod_{i=1}^n dt_i~e^{-\frac{T_n}{\tau_a}}
\mathcal{F}_n(t,\{T_j\})x_r(t-T_n),\nonumber 
\end{eqnarray}
where $\tau_d = \gamma/k_d$ is the equally weighted difference time
constant and
\begin{eqnarray}
T_n &\equiv&  t_1 + t_2 + t_3 + \cdots + t_n,\\
\mathcal{F}_n(t,\{T_j\}) &\equiv& f(t-T_1)f(t-T_2) \cdots f(t-T_n);
\end{eqnarray}
and the following conventions are used:
\begin{equation}
\mathcal{F}_0(t,\{T_j\}) \equiv 1\nonumber
\end{equation}
and for arbitrary function $H(t,\{T_j\})$
\begin{equation}
\int_0^\infty \prod_{i=1}^0 dt_i H(t,\{T_j\}) \equiv H(t).\nonumber
\end{equation}

Following these conventions the zeroth order term in
\eref{eqn:xt} is simply $x_r(t)$, which is just the thermal
motion.  Addition of the next term provides an approximation for the
solution, whose form is
\begin{equation}
x(t) \approx x_r(t) + \frac{1}{\gamma}\int_0^\infty dt'
e^{-t'/\tau_a}f(t-t')k_dx_r(t-t').
\end{equation}
This solution has the physical interpretation that the motion of the
probe is a linear combination of the thermal motion $x_r(t)$ plus the
thermal motion subjected to the fluctuating spring force
$-f(t)k_dx_r(t)$.  The higher order corrections describe correlations
which impose self-consistency for $x(t)$.  Truncation at second order
would be reasonable if thermal motion were dominate over switching
motion, $x_r(t) \gg x_f(t)$.  In general, however, we are {\em not}
interested in such cases, because extraction of the molecular
properties from experiments becomes quite difficult when thermal noise
dominates over switching.  Moreover, we find it more useful to expand
the auto-correlation function around two physically important,
limiting cases.

Using \eref{eqn:xt} a series solution for the auto-correlation
function is obtained:
\begin{widetext}
\begin{equation}
\av{\delta x(t)\delta x(0)} = \label{eqn:xx}
\sum_{n,n'=0}^\infty
\left(\frac{1}{\tau_d}\right)^{n+n'}\int_0^\infty\prod_{i=1}^n\prod_{i'=1}^{n'}dt_idt_{i'}e^{-\frac{T_n+T_{n'}}{\tau_a}}
\av{\mathcal{F}_n(0,\{T_j\})\mathcal{F}_{n'}(t,\{T_{j'}\})x_r(-T_n)x_r(t-T_{n'})}.
\end{equation}
\end{widetext}
Using the fact that within our approximations the fluctuating spring
constant is statistically independent of the thermal force,
\eref{eqn:decouple}, dictates that the motion of the probe induced by
thermal fluctuations $x_r(t)$ is independent of the switching function
$f(t)$.  The averages over the thermal induced displacements $x_r(t)$
and the stochastic variable $f(t)$ then decouple,
\begin{eqnarray}
\av{f(t_1)\cdots f(t_m)x_r(t)x_r(t')} &=& \av{f(t_1)\cdots
f(t_m)}\av{x_r(t)x_r(t')} \nonumber\\
&=&\frac{\kT}{k_a}e^{-\frac{|t-t'|}{\tau_a}}\av{f(t_1)\cdots
f(t_m)}.\nonumber
\end{eqnarray}

The auto-correlation function may now be written as
\begin{widetext}
\begin{equation}
\av{\delta x(t)\delta x(0)} = \label{eqn:xx2}
\frac{\kT}{k_a}\sum_{n,n'=0}^\infty
\left(\frac{1}{\tau_d}\right)^{n+n'}\int_0^\infty\prod_{i=1}^n\prod_{i'=1}^{n'}dt_idt_{i'}
\av{\mathcal{F}_n(0,\{T_j\})\mathcal{F}_{n'}(t,\{T_{j'}\})}e^{-\frac{|t+T_n-T_{n'}|+T_n+T_{n'}}{\tau_a}},
\end{equation}
\end{widetext}
where the stochastic features regarding the switching of the molecule
is embedded in the term
$\langle\mathcal{F}_n(0,\{T_j\})\mathcal{F}_{n'}(t,\{T_{j'}\})\rangle$.
Although \eref{eqn:xx2} provides a valid series for the general
solution for the auto-correlation function, we find it more
illuminating and beneficial to begin such a series from the two
obtained bounds found for the auto-correlation function.  We now
derive (mathematically, as opposed to physically) the solutions for
the auto-correlation function in the fast- and slow-switching limits
directly from \eref{eqn:xx2}.  This derivation will allows us to
express the general solution for the auto-correlation function in the
terms of two series solutions initiated at these limiting forms.

\subsection{The fast- and slow-switching limits} \label{sec:limits}

In our previous analysis, \sref{sec:numac}, we introduced two
possible solutions to the auto-correlation function based on physical
arguments and proved in \sref{sec:theory} that these two
solutions provided an upper and lower bound to the general solution
for the auto-correlation function.  We denoted these two limits as the
fast-switching limit and the slow-switching limit.  In this section we
derive these solutions starting from the series solution expressed in
\eref{eqn:xx2}.

Mathematically, the fast-switching limit is defined as the case when
$\kaps \gg 1/\tau_a$, where $\kaps$ is the {\em slowest} rate of
transition between the different states ($\kaps = \min(\kappa_{ij})$
for $i \neq j$).  This states that the molecule switches between
states on times scales that are much shorter than those governing the
motion of the probe.  The slow-switching limit is defined as the case
when $\kapf \ll 1/\tau_a$, where $\kapf$ is the {\em fastest} rate of
transition between different states ($\kapf = \max(\kappa_{ij})$ for
$i \neq j$).  This states that the time scales governing the motion of
the probe is much shorter than those governing the transitions of the
molecule.  For the problems of interest $\tau_a$ (where $\tau_a =
\gamma/(k_{\rm P}+(k_{\rm m}^{\rm max}+k_{\rm m}^{\rm min})/2)$ and
$k_{\rm m}^{\rm max} = \max(k_{{\rm m}i})$ and $k_{\rm m}^{\rm min} =
\min(k_{{\rm m}i})$) can be considered as a measure of the correlation
time of the experimental probe, as the compliance of the probe
($1/k_{\rm P}$) is generally much smaller or of order of the
compliance of the biomolecule ($1/k_{{\rm m}i}$) in any state.  (See
discussion in \aref{app:tc}.)

The auto-correlation function, \eref{eqn:xx2}, contains averages over
the random variable $f(t)$ along with a decaying exponential with
time constant $\tau_a$.  In the fast-switching limit the averages over
the random variable $f(t)$ decouple on time scales much faster than
any other in the problem.  Therefore, we can assume that any
correlation between the random variables ($f(t)$) at different times
are negligible and make the following approximation:
\begin{eqnarray}
&& \av{\mathcal{F}_n(0,\{T_j\})\mathcal{F}_{n'}(t,\{T_{j'}\})} =
\label{eqn:Ffs} \\ &&\av{f(-T_1)\cdots f(-T_n)f(t-T_{1'})\cdots
f(t-T_{n'})} \approx
\av{f}^{n+n'},  \nonumber
\end{eqnarray}
where
\begin{equation}
\av{f}^m = \left(\sum_i f_i P_i \right)^m.
\label{eqn:Cinfty}
\end{equation}
Here $f_i$ is the value of $f$ when in state $i$, \eref{eqn:fi},
and we have used the stationary condition for $f(t)$,
\eref{eqn:statcond}.

In the slow-switching limit the averages over the random variable
$f(t)$ will remain nearly constant on time scales where the integrand
is non-negligible.  Therefore, we can assume that the averages have no
dependence on time differences and we make the following approximation
\begin{eqnarray}
 && \av{\mathcal{F}_n(0,\{T_j\})\mathcal{F}_{n'}(t,\{T_{j'}\})} = \label{eqn:Fss} \\ &&\av{f(-T_1)\cdots
f(-T_n)f(t-T_{1'})\cdots f(t-T_{n'})}  \approx\av{f^{n+n'}}, \nonumber
\end{eqnarray}
where
\begin{equation}
\av{f^{m}} = \sum_i P_i f_i^m.\label{eqn:C0}
\end{equation}

An important result is the fact that in these limits the averages over
the random variable are independent of any time variables.  It is
shown in \aref{app:avC} that if
\begin{equation}
\av{f(t_1)f(t_2)...f(t_m)} = C_m,\label{eqn:Cn}
\end{equation}
where $C_m$ is independent of the time variables, then the integrals
in the series (\eref{eqn:xx2}) can be done explicitly and the
solution for the auto-correlation function may be expressed as
\begin{equation}
\av{\delta x(t) \delta x(0)} =
\frac{kT}{k_a}e^{-t/\tau_a}\sum_{m=0}^\infty
\left(\frac{\tau_a}{\tau_d}\right)^m C_m\sum_{j=0}^m
 \frac{1}{j!}\left(\frac{t}{\tau_a}\right)^j. \label{eqn:xxC}
\end{equation}
For the two limiting cases
(\erefs{eqn:Cinfty}~and~\ref{eqn:C0}) this result can be
summed in closed form and the resulting auto-correlation functions
are:
\begin{equation}
\av{\delta x(t)\delta x(0)} = \left\{ \begin{array}{lll}
\frac{\kT}{\langle k\rangle}e^{-t \avtauinv} & & \kaps \gg 1/\tau_a,
\\ &&\\ \left\langle \frac{\kT}{k}e^{-t/\tau}\right\rangle& &\kapf \ll
1/\tau_a.
\end{array} \right. \label{eqn:kt} 
\end{equation} 
These are precisely the physically motivated solutions we have
presented in the previous sections.   

As expressed in \sref{sec:numac}, the rate of switching
influences the effective interaction between the biomolecule and
experimental probe.
In the fast-switching limit the biomolecule switches between states so
``fast'' that it equilibrates on the time scales shorter than those
governing the motion of the probe (conveyed mathematically in
\eref{eqn:Ffs}) and; therefore, the probe ``sees'' the
biomolecule as a stable single-state system with an effective spring
constant equal to the average spring constant of the biomolecule
$\av{k_{\rm m}}$,
\begin{equation}
\av{k} = \sum_i P_ik_i = k_\text{P} + \av{k_{\rm m}}. \label{eqn:kave}
\end{equation}
In this case the motion of the probe cannot be used to delineate the
different states, but only infer an effective interaction.  In the
slow-switching limit thermal decay causes the experimental probe to
loose memory of any previous position on times scales that are much
smaller than the lifetime of each state (conveyed mathematically in
\eref{eqn:Fss}) and; therefore, the probe is able to fully resolve
each states auto-correlation function, properly weighted.  It is
reassuring to us that these limiting cases appeal to physical
intuition: in the fast-switching limit it is the average spring
constant that is physically observed, while in the slow-switching
limit it is the average auto-correlation function that is physically
observed.

\subsection{Series expansion around limiting forms} \label{sec:limiting}

Armed with the derivation of the two limiting forms from our general
series expression, we now express the general solution for the
auto-correlation function in the form a series initiated from either
of the limiting forms.  The deviation of the auto-correlation function
from the limiting forms reveals the time scales on which the
biomolecule fluctuates between various states, as was illustrated in
\sref{sec:numac} (\frefs{fig:ac_num}~and~\ref{fig:ac_num_largek1}).
To obtain the two new general series solutions we now add to the
general solution, \eref{eqn:xx2}, the limiting forms, \eref{eqn:kt},
and subtract from each term the contribution that is already captured
in that limiting form (\eref{eqn:Cinfty}~or~\ref{eqn:C0}).

Initiated at the fast-switching limit the general solution for the
auto-correlation function takes the form
\begin{widetext}
\begin{equation}
\av{\delta x(t)\delta x(0)} = \frac{\kT}{\av{k}}e^{-t\avtauinv} +
\label{eqn:xx2inf}  \frac{\kT}{k_a}\sum_{n,n'=0}^\infty
\left(\frac{1}{\tau_d}\right)^{n+n'}
\int_0^\infty\prod_{i=1}^n\prod_{i'=1}^{n'}dt_idt_j 
\Di{n,n'}(t,\{T_j\},\{T_{j'}\}) e^{-\frac{|t+T_n-T_{n'}|+T_n + T_{n'}}{\tau_a}},
\end{equation}
\end{widetext}
where
$$ \Di{n,n'}(t,\{T_j\},\{T_{j'}\}) \equiv
\av{\mathcal{F}_n(0,\{T_j\})\mathcal{F}_{n'}(t,\{T_{j'}\})} - \Ci{n+n'}
$$ and the latter term ($\Ci{n+n'}$) is the contribution to the
limiting form.  Here the superscript ($\infty$) denotes that the
expansion is around the fast-switching limit ($\kaps\tau_a~\gg~1$).
Initiated at the slow-switching limit the general solution for the
auto-correlation function takes the form
\begin{widetext}
\begin{equation}
\av{\delta x(t)\delta x(0)} =
\left\langle\frac{\kT}{k}e^{-t/\tau}\right\rangle +
\label{eqn:xx20}  \frac{\kT}{k_a}\sum_{n,n'=0}^\infty
\left(\frac{1}{\tau_d}\right)^{n+n'}
\int_0^\infty\prod_{i=1}^n\prod_{i'=1}^{n'}dt_idt_j
\Dz{n,n'}(t,\{T_j\},\{T_{j'}\}) e^{-\frac{|t+T_n-T_{n'}|+T_n+T_{n'}}{\tau_a}},
\end{equation}
\end{widetext}
where
$$ \Dz{n,n'}(t,\{T_j\},\{T_{j'}\}) \equiv
\av{\mathcal{F}_n(0,\{T_j\})\mathcal{F}_{n'}(t,\{T_{j'}\})} - \Cz{n+n'}
$$ and the latter term ($\Cz{n+n'}$) is the contribution to the
limiting form.  Here the superscript (0) denotes that the expansion is
around the slow-switching limit ($\kapf\tau_a \ll 1$).  Note, that by
definition
\begin{equation}
\Di{0,0} = \Dz{0,0} = \Di{0,1} = \Di{1,0} = \Dz{0,1} = \Dz{1,0} =
0 \label{eqn:Di1Dz1}
\end{equation}
and the leading order correction is second order in both series.

The two series, \erefs{eqn:xx2inf}~and~\ref{eqn:xx20}, are general
solutions to the auto-correlation function which initiate at opposing
limits and approach the solution from opposite directions, adding to
utility of expanding the solution around these two limiting forms.
The expansion terms in \eref{eqn:xx2inf} must add up to a positive
number and those in \eref{eqn:xx20} must add up to a negative number.

When the solution for the auto-correlation function corresponds to one
of the limiting forms the auto-correlation function cannot be used to
convey the rates of transition, as these limiting solutions only
depend on probabilities.  The magnitude of the rates of transitions
are conveyed through the terms $\Di{n,n'}$ and $\Dz{n,n'}$.  When the
time dependence for these terms reveal themselves through the solution
to the auto-correlation function can the magnitude of the rates of
transition be determined.  This time dependence will be revealed if
these terms ($\Di{n,n'}$ and $\Dz{n,n'}$) vary on the time scale
$\tau_a$.  We therefore, expect that measurements of the
auto-correlation function would be most applicable to determine the
rates of transitions when the rates are of order of the inverse time
constant of the experimental probe, $\{\kappa_{ij}\} \approx
1/\tau_a$, for $i \neq j$ (not near either of the limiting forms as
exemplified in \frefs{fig:ac_num}~and~\ref{fig:ac_num_largek1}).  This
is a time resolution that is inaccessible by traditional analysis such
as that presented in \sref{sec:background}.

Although we believe that this method will be most amiable to
experimental analysis when the solution for the auto-correlation
function is not near one of the limiting forms, analysis of the
auto-correlation function may still provide valuable information
regarding the rates of transition when this condition does not hold.
For the case when the system is in the fast-switching limit, analysis
of the auto-correlation function does not provide enough temporal
resolution to resolve the absolute rates of transition, the probe
``sees'' the average spring constant; however, if the system is in the
fast-switching limit, analysis of the auto-correlation function
provides a lower bound on the rates, as by definition $\{\kappa_{ij}\}
\gg 1/\tau_a$ for $i \neq j$.  This lower bound is far greater than
that which can be concluded by traditional methods, as that presented
in \sref{sec:background} (see \frefs{fig:average}~(c)~and~(d)).  When
the solution for the auto-correlation function is properly represented
by the slow-switching limit traditional methods like that presented in
\sref{sec:background} are applicable and far superior.

If the calculation of the auto-correlation function from the series
expansion (\eref{eqn:xx2inf}~or~\eref{eqn:xx20}) and its comparison
with experimental results is practical, the synopsis presented here
supports the notion that the calculation of the auto-correlation
function may be used as method to determine the rates of transitions
for macromolecular systems, those which are inaccessible with current
analytic methods.  To help garner insight into the applicability of
this method we now explore issues regarding the convergence of both
series to the general solution.

\section{Convergence and Applicability} \label{sec:convergence}

A key to the applicability of this method will be on the ease at which
the series expansions converges to the general solution; in this
section we present an analysis regarding this convergence.  We then
offer a concrete understanding of how the biomolecular properties
affects both the convergence of each series onto the general solution
and the applicability of this method to study rates of transition by
focusing on the special case of the two state problem.  The two state
problem is further explored, expressions for each series expansion (up
to third order) are presented for a Markov process, expressions which
are applicable to experimental analysis.  Guided by the insight
obtained in the section, regarding the convergence of each series to
the general solution, we then test these expressions to predict the
form of the auto-correlation functions calculated numerically in
\sref{sec:numac}.

We now generate expressions which lend insight into the convergence of
each series.  We portray the convergence of both series through the
analysis of a new series, which we term as the difference series.
This difference series measures the separation between both series
solution when truncated at a given order.  The utility of analyzing
the difference series is that it can be written in a simplified
closed form and; therefore, readily studied.

Although this difference series does not provide the error associated
with truncating a particular series, it stills proves to be a
practical measure for the convergence of each series onto the general
solution.  This is because both series start at opposing bounds and
must converge to the same result.  In addition the utility of this
methodology for determining the rates of transition will depend on the
condition that the solution for the auto-correlation function is
separated from both limiting forms and; therefore, ``equally''
separated from both series.  The separation between both series at a
given order should then serve as a reasonable upper-bound
estimate~\footnote{We do not consider the pathological case where the
values of both series are nearly equivalent beyond some point in the
expansion, however neither individual series have converged to the
solution with the same accuracy.} for the separation of an individual
series from the true solution.

We define the difference series as
\begin{equation}
\Delta A_m(t) \equiv \av{\delta x(t)\delta x(0)}^{(0)}_m - \av{\delta
x(t)\delta x(0)}^{(\infty)}_m, \label{eqn:Amt_d}
\end{equation}
where $\Delta A_m(t)$ is the difference between both series when they
are truncated at order $m$.  The superscript on the correlation
functions denotes the particular series from which the
auto-correlation function is inferred and the subscript denotes the
order of truncation.  This difference, at the $m^{th}$-order, is
\begin{widetext}
\begin{eqnarray}
\frac{\Delta A_m(t)}{\kT} &=& \left\langle\frac{1}{k}e^{-t/\tau}\right\rangle -
\frac{1}{\av{k}}e^{-t\avtauinv}
 + \frac{1}{k_a} \sum_{n,n'=0}^{n+n'=m}
\left(\frac{1}{\tau_d}\right)^{n+n'}
\int_0^\infty\prod_{i=1}^n\prod_{i'=1}^{n'}dt_idt_je^{-\frac{t_i+t_{i'}}{\tau_a}}
\bigl(\Ci{n+n'} - \Cz{n+n'}\bigr)
e^{-\frac{|t+T_n-T_{n'}|}{\tau_a}},\nonumber
\end{eqnarray}
\end{widetext}
where the sum is restricted to $n+n'\leq m$. Noting that
$\bigl(\Ci{n+n'} - \Cz{n+n'}\bigr)$ is independent of time we make use
of \eref{eqn:xxC} and rewrite this difference as
\begin{eqnarray}
\frac{\Delta A_m(t)}{\kT} &=& \label{eqn:Amt} 
\left\langle\frac{1}{k}e^{-t/\tau}\right\rangle -
\frac{1}{\av{k}}e^{-t\avtauinv} +
\frac{1}{k_a}e^{-t/\tau_a}\\ && \times\sum_{n=0}^m
\left(\frac{\tau_a}{\tau_d}\right)^n \bigl(\Ci{n} - \Cz{n}\bigr)
\sum_{j=0}^n \frac{\bigl(t/\tau_a\bigr)^j}{j!}.  \nonumber
\end{eqnarray}
The difference series is now expressed in a simplified form, more
amiable to analysis.  The difference series can be further simplified
for the $t = 0$ case, where the difference series can be summed into a
closed form
\begin{equation}
\frac{\Delta A_m(0)}{\kT} =  \label{eqn:Am0} 
\left\langle\frac{\bigl(1 - k/k_a\bigr)^{m+1}}{k} \right\rangle -
\frac{\bigl(1-\av{k}/k_a\bigr)^{m+1}}{\av{k}}.
\end{equation}
The advantage of exploring the convergence of the difference series
$\Delta A_m(t)$, as oppose to each individual series, is that it is
much simpler to evaluate.  The difference series only depends on
equilibrium averages ($\langle f^m\rangle$) instead of integrals
containing correlations ($\Di{n,n'}, \Dz{n,n'}$).

For large times the difference series will be dominated by the
decaying exponentials, even though the latter
summation~\cite{brissaud74} ($j$-summation in \eref{eqn:Amt})
increases with time.  Therefore, the magnitude of the difference
series will be dominated by short times, $t < \tau_a$.  Because of
this we use the simple $t = 0$ difference, \eref{eqn:Am0}, as a
measure of the error in either series; which is more amiable to
analysis.

As an aside, we note that if expansion terms are calculated for one
series expansion then the corresponding terms in the other series
expansion can simply be calculated through the employment of
\erefs{eqn:Amt_d}~and~\ref{eqn:Amt}.  This can greatly reduce
the work in calculating the second series.

Using the difference series as our guide, we now discuss how the
convergence of the series depends on various parameters in the problem
and determine a range for which this method may be applicable for the
analysis of experimental data.  For concreteness we focus on the two
state problem, where biomolecules exist in two distinct mechanical
states.  We envision an experimental situation similar in fashion to
that in \frefs{fig:example}~and~\ref{fig:ob} (except where the
macromolecule can only access two states).  The two state problem is
both revealing and serves as the basis of a number of actual
applications of these ideas.  In addition to determining a range of
parameters for which this method may be applicable we present
truncated series solutions (up to third order) for the two state
Markov process; giving functional form to the solution to the
auto-correlation function when it deviates from the limiting
solutions.

\subsection{The two state problem}

One of the key ambitions of this section is to determine a range of
biomolecular parameters for which the analysis of the auto-correlation
function will be beneficial for revealing such properties,
particularly the rates of transition.  As noted, the form of the
auto-correlation function depends on the rates of transition more
sensitively when the rates are of order of the inverse of the time
constant of the experimental probe, $\{\kappa_{ij}\} \approx 1/\tau_a,
i\neq j$.  Under such conditions the form of the auto-correlation
function is influenced by the magnitude of the rates of transition and
not just the probabilities of lying in a particular state.  Therefore,
a criterion for applicability is that the rates of transition are of
order of the inverse time constant of the experimental probe.

However, satisfying this criterion alone will not guarantee the
applicability of this method.  For an experimental study to exploit
this method it must be able to attain sufficient spatial resolution to
differentiate between various possible solutions, that is it must be
capable of discerning between solutions in the fast-switching limit,
the slow-switching limit or somewhere in between.  The spatial
resolution is directly correlated with the separation of the two
bounding solutions, the fast- and slow-switching limits; a smaller
separation requires greater spatial capabilities of the experimental
apparatus.  As we have demonstrated in \sref{sec:numac} the separation
between the slow-switching and fast-switching limits is affected by
the ratio of the spring constants between the two states, as
exemplified in \frefs{fig:ac_num}~and~\ref{fig:ac_num_largek1}.
Imposing applicability will impose a restriction regarding the ratio
of the spring constants of the states.  Additionally, we will show
that applicability will enforce limits on the probability of the
system to lie within a given state.  This later enforcement is simple
understood from the fact that if one state dominates, the influence of
the other will not appreciably affect the form of the auto-correlation
function and; therefore, difficult to detect.  Finally, for this
analysis to determine the rates of transition from experimental
measurements of the auto-correlation function requires terms within
the series to be computed.  Calculations of such terms can be tedious,
leading to the final constraint that applicability requires adequate
convergence to occur within a limited number of expansion terms.  This
last condition combined with the need for deviation from the limiting
forms restricts the parameter space for which this method should be
applicable.

We have given a number of conditions, which will restrict the
parameter space, for when we believe this method will be applicable to
experimental analysis.  To determine this space of molecular
parameters we now proceed by going through a number of refinement
procedures.  The first two conditions we explore are concerned with
the requirement that to amiable to experimental analysis (for
determining the rates of transition) requires the solution for the
auto-correlation function to be ``well'' separated from both limiting
forms.  This will entail a condition regarding the probability of the
system to lie within a given state and a condition regarding the ratio
of the spring constants.  Finally, in order to limit the number of
integral evaluations (number of terms in the series) required for a
desired convergence, an additional condition regarding the spring
constant ratio is imposed.

\subsubsection{Parameter space of applicability for the two state problem}

As mentioned at the end of \sref{sec:limiting}, the auto-correlation
function can in principle be used to determine the individual rates of
the macromolecule when it differs from both limiting forms
($\av{\delta x(t)\delta x(0)}^{(\infty)}_0$ or $\av{\delta x(t)\delta
x(0)}^{(0)}_0$).  Because both limiting forms provide opposing bounds
to the solution for the auto-correlation function, a necessary
condition for applicability requires the bounds to differs from each
other more than the available spatial capabilities of the experimental
apparatus.  Therefore, to obtain an initial range of parameters for
which this method may be of use, we study the difference series at $t
= 0$ and $m=0$ ($\Delta A_0(0)$), which serves as a measure of
``nearness'' for both limiting forms.  ($\Delta A_0(0)$ is the
difference between the limiting forms.)  When $\Delta A_0(0)$ is large
then the limiting forms are well separated and the solution for the
auto-correlation function is not ``near'' the limiting forms. (Recall,
we are considering cases for which $\{\kappa_{ij}\} \approx 1/\tau_a,
i\neq j$ and; therefore, the solutions lies in between the limiting
forms.)

Prior to proceeding with this analysis, we need a definition of
``nearness''.  For concreteness we consider an experiment whose error
tolerance for calculating the auto-correlation function is of the
order of 5\% of the mean-squared displacement of the experimental
probe $\av{x^2}$.  To obtain a range of parameters we seek situations
where both limiting forms differ by more than 5\%.

\begin{figure}
\begin{center}
\rotatebox{90}{\hspace*{1.25cm} $\frac{\Delta A_0(0)}{\langle
X^2\rangle_\infty } = \av{k}\left\langle \frac{1}{k}\right\rangle -1$}
\scalebox{0.4}{\includegraphics{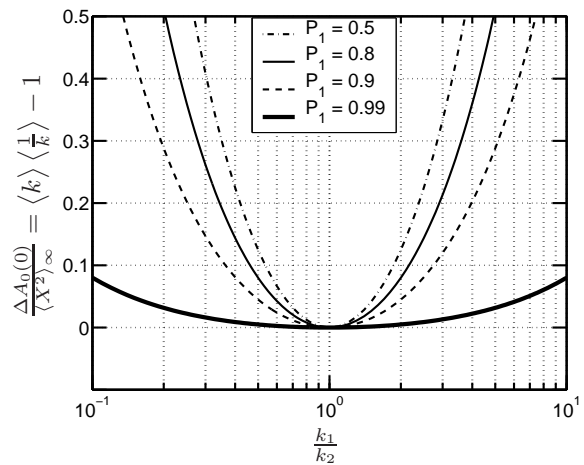}} \\ \ \ \ \ \ \
$\frac{k_1}{k_2}$
\caption{Difference series ($t = 0, m=0$) $\Delta A_0(0)$ normalized
by the solution in the fast-switching limit $\langle
x^2\rangle_\infty$ as a function of the spring constant ratio, at
various probabilities. (P$_1$ is the probability of being in state 1.)
Note that when this ratio is 1 both statse are mechanically
indistinguishable.}
\label{fig:xx0dxx0}
\end{center}
\end{figure}

Figure~\ref{fig:xx0dxx0} displays the difference series at $t = 0$ and
$m=0$, normalized by the solution in fast-switching limit. (The
fast-switching limit is an approximate measure of $\av{x^2}$.)  The
results are plotted as a function of the spring constant ratio, at
various probabilities.  The limiting forms give similar answers
($\Delta A_0(0)$ is small) if one state is dominate (very large
probability), so long as the spring constants do not vary by many
orders of magnitude.  Here, we only consider experimental situations
where the spring constants in both states are of the same order of
magnitude.  Cases for which this differs are unlikely for the
experimental situations of interest.~\footnote{A situation where the
spring constant ratio would differ by orders of magnitude would occur
when the spring constant for the two biomolecular states differ by
orders of magnitude and the larger of the two biomolecular spring
constants is orders of magnitude larger than the intrinsic spring
constant of the experimental probe. Such cases are generally unlikely
and are not consider here.}  Therefore, to consider cases where the
difference in the bounds is greater than 5\%, a necessary condition
that we impose is the probability of being in either state must be
greater than 10\%.

Having produced an initial bound on the probabilities for
applicability of this analysis, we now seek an initial lower bound on
the ratio of the spring constants.  For argument sake we now choose
state 1 to be defined such that $k_1 > k_2$.  When $k_1 \approx k_2$,
either the mechanical properties of the macromolecular states are
similar or the spring constant of the experimental probe is dominate
over that of the molecule.  In such cases the $t = 0$ solutions for
both bounds are nearly identical and it should become quite difficult
to use the general solution for the auto-correlation function to
extract the rates of transition.  For the bounds to have a difference
greater than 5\% (our predefined tolerance) then the spring constant
ratio ($k_1/k_2$) should be greater than 1.5.  This puts a limitation
on the stiffness of the experimental probe as compared to that of the
macromolecule.  For the spring constant ratio to be greater than 1.5,
then the macromolecular stiffness in state 1 must be greater than 1/2
the stiffness of the experimental probe.

\begin{figure}
\begin{center}
 (a) \\ \rotatebox{90}{\hspace*{1.5cm}$A_m(0)\av{k}/\kT$}
 \scalebox{0.4}{\includegraphics{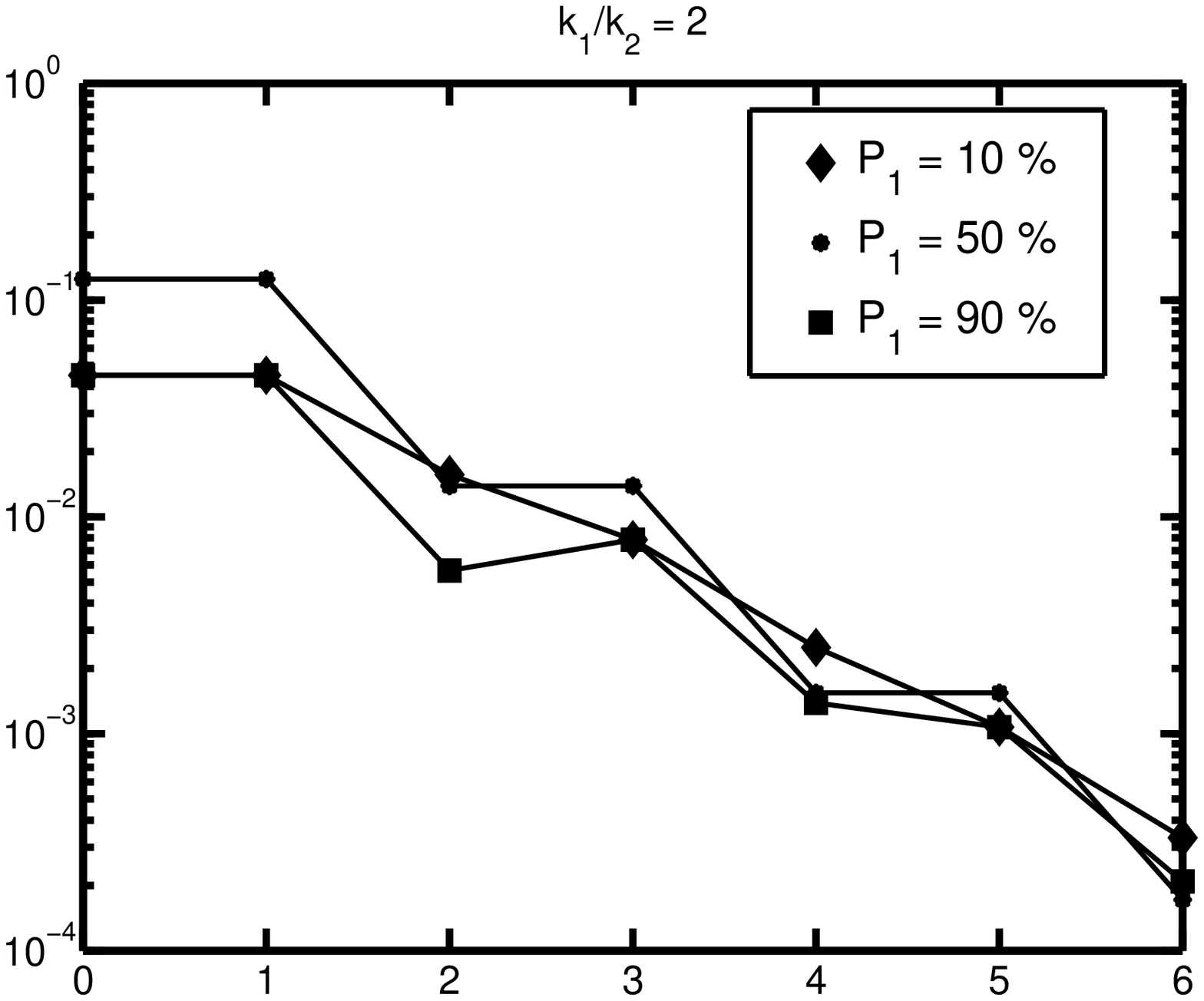}} \\ Number of Expansion
 Terms \\
\vspace{.2cm} (b) \\
 \rotatebox{90}{\hspace*{1.5cm} $A_m(0)\av{k}/\kT$}
\scalebox{0.4}{\includegraphics{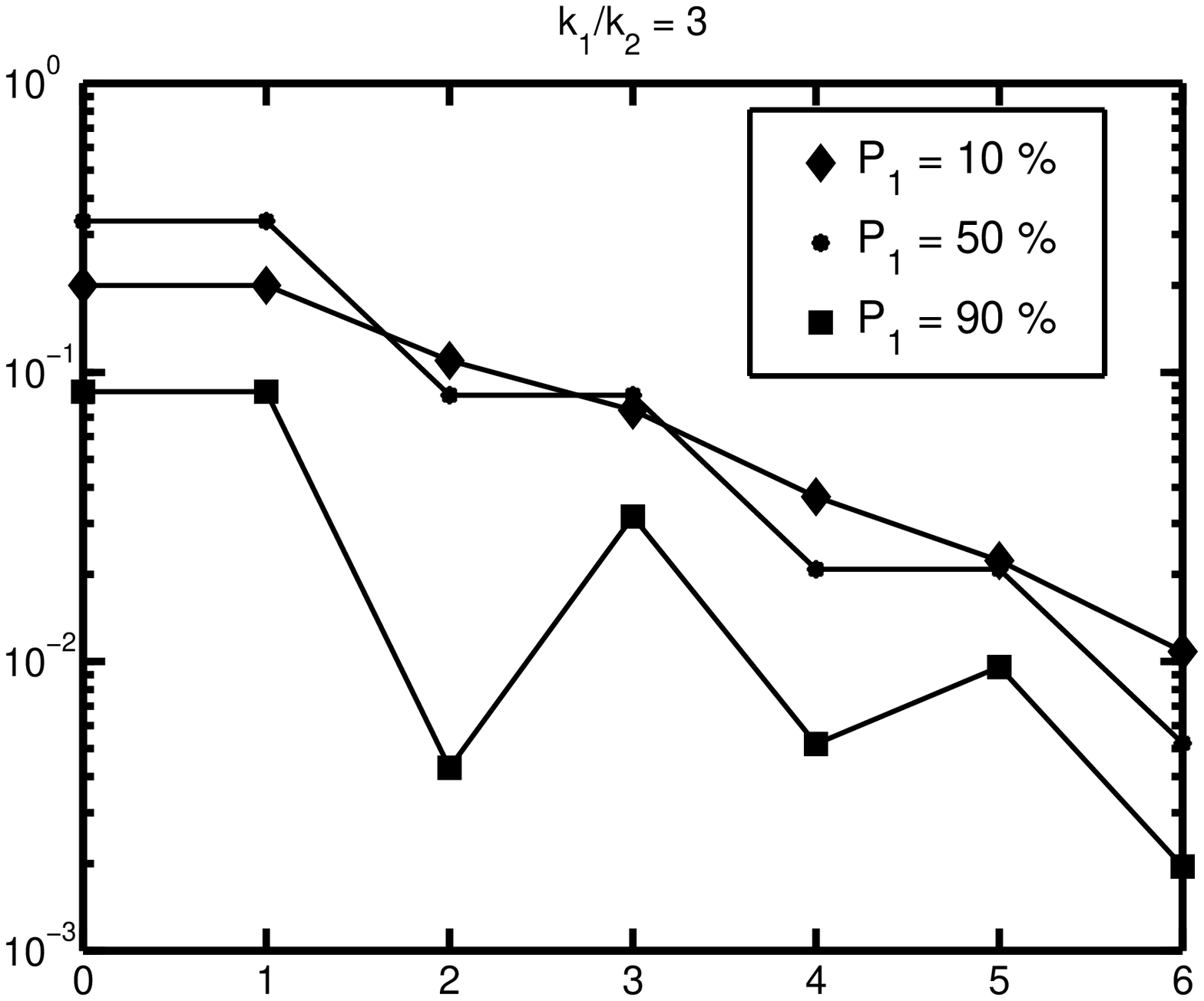}} \\  Number of
Expansion Terms 
\caption{Difference series $A_m(0)$ normalized by $\av{k}/\kT$ plotted
as a function of the number of expansion terms, for $k_1/k_2 = 2$ (a)
and $k_1/k_2 = 3$ (b), for various probabilities. P$_1$ is the
probability of lying in state 1.}
\label{fig:deltaA2}
\end{center}
\end{figure}

Satisfying these two conditions, the solution for the auto-correlation
function deviates from the limiting forms on the order of 5\% or
greater.  However, this does not guarantee applicability.  The next
condition that we impose regards limiting the number of terms in the
series needed for adequately convergence to the desired solution.
Here we are interested in cases for which the series solutions,
\erefs{eqn:xx2inf}~and~\ref{eqn:xx20}, are within the predefined
tolerance (5\% of $\av{x^2}$) of the exact solution only after the
calculation of three expansion terms.  \Fref{fig:deltaA2} plots the $t
= 0$ difference series $\Delta A_m(0)$ (normalized by $\kT$/$k_a$) for
two different spring constant ratios at various probabilities, as a
function of the number of expansions terms.  We choose the spring
constant ratio to be 2 in figure (a) and 3 in figure (b), recalling
that it must be greater than 1.5 to be within our initial refined
parameter list.  From \fref{fig:deltaA2}(a) it is evident that
truncation of the series solution at second order is sufficient to
allow for convergence to within 5\% accuracy of the exact solution.
From \fref{fig:deltaA2}(b), however, it takes more terms.  To be
within the experimental tolerance of 5\%, the series must contain at
least four terms.

As stated in \sref{sec:theory} (\eref{eqn:Di1Dz1}) there
are no first order corrections; therefore, only one (three) expansion
term (terms) are needed to be within this predefined error tolerance
for these cases.  For spring constants ratios greater than 3 more
terms are needed.  Therefore, this analysis gives an upper bound on
our spring constant ratio to be less than 3.

The analysis presented in this section gives a range of parameters for
which we believe this theory should be useful and practical for
determining the rates of transition (for the two state problem) from
experimental measurements, those which are not accessible from
traditional means.  To determine the rates of transition the solution
for the auto-correlation function must deviate from both limiting
forms, requiring the rate of transitions to be of the order of the
inverse time constant of the experimental probe, $\{\kappa_{ij}\}
\approx 1/\tau_a, i \neq j$.  To be within 5\% accuracy of the exact
solution and not near one of the limiting forms, the probability of
being in either state should be greater than 10\% and the ratio of the
spring constants should also be greater than 1.5.  To limit the number
of calculated terms in the series to three or less, the ratio of the
spring constants should also be less than~3.

This range of parameters is somewhat limited; however, this method
still provides an alternative to standard analysis, expanding the
temporal capabilities alloted in experimental studies.  Having
obtained this range of parameters for applicability we now present
calculated terms in both series, up to third order, for the two state
Markov process.  These solutions are amiable to experimental analysis,
whose form depends on the value of the individual rates.  Guided with
our understanding of convergence obtained in this section, we then
compare our truncated analytic series solutions to the numerical
results presented in \sref{sec:numac}.

\subsubsection{Solutions for the Markov process}

Our system of interest is a two state Markov process, a system
consisting of two distinct mechanical states (different spring
constants) whose transitions are local in time (independent of all
previous configurations).  \aref{app:markov} presents the detailed
mathematics describing the two state Markov process and the solution
for the auto-correlation function truncated at third order,
\erefs{eqn:thirdorder}~and~\ref{eqn:thirdorder0}.  Here we compare
those solutions with the numerical solutions presented in
\sref{sec:numac} and determine the degree of convergence.

\begin{figure}
\begin{center}
\centerline{\epsfxsize=3.0truein \epsfbox{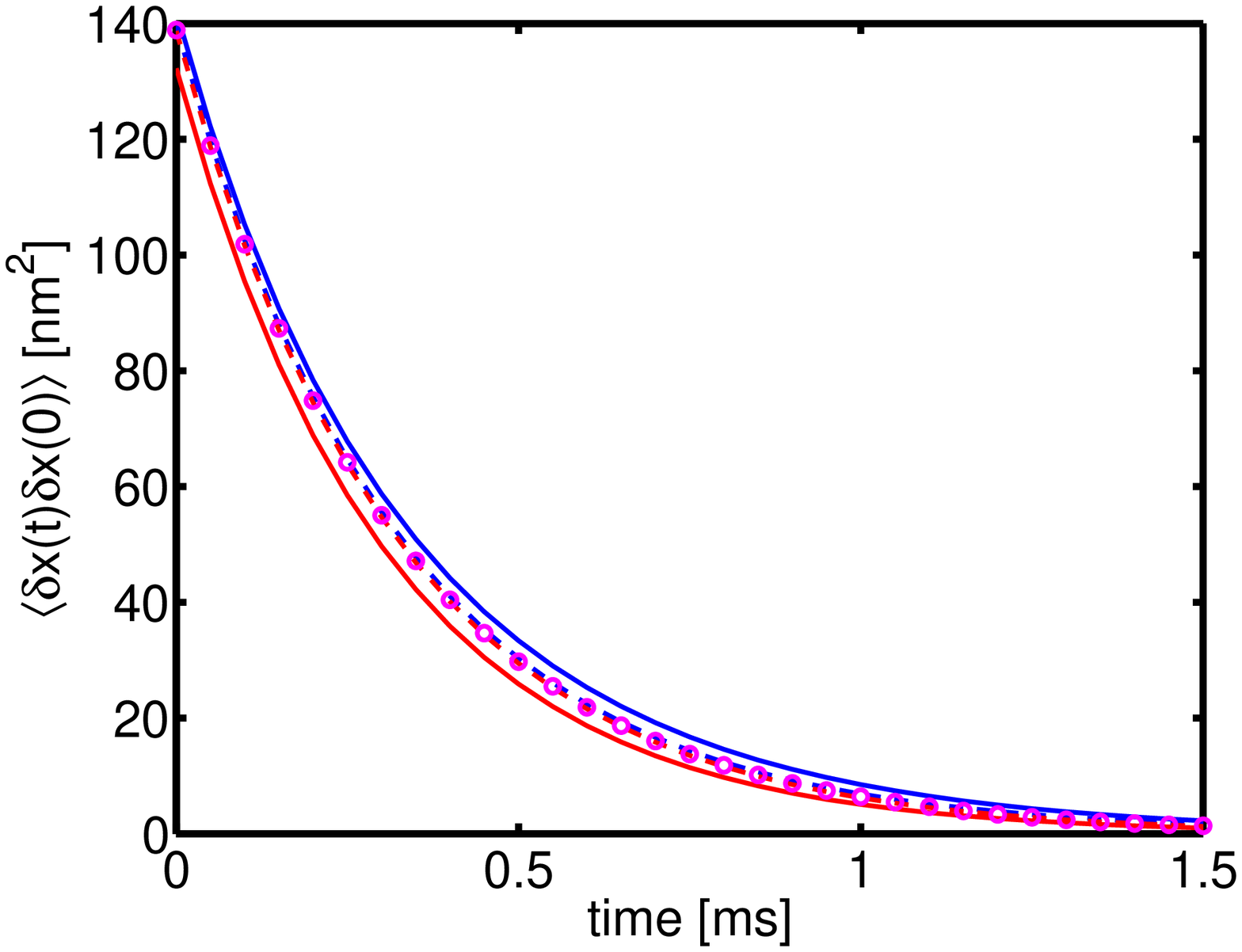}} 
\caption{Auto-correlation function for a system with a fluctuating
spring constant.  Numerical solutions for case~D is represented by the
red circles. The solid red line is the solution in the fast-switching
limit and the blue line is the solution in the slow switching limit.
Series solutions for the auto-correlation function, truncated at third
order, are represented by the red dashed-line when initiated from the
fast-switching limit solution (\eref{eqn:thirdorder}) and by the
blue dashed-line when initiated from the slow-switching limit solution
(\eref{eqn:thirdorder0}).}\label{fig:ac_pert}
\end{center}
\end{figure}

\Fref{fig:ac_pert} redisplays the numerical solutions for case~D
presented in \sref{sec:numac}.  Recall, case~D corresponds to a system
where the rates of transition are of order of the inverse of the
intrinsic time constant of the experimental probe (see
Table~\ref{tab:rates}).  In addition, solutions corresponding to the
fast-switching limit (solid blue line) and slow-switching limit (solid
red line), see \fref{fig:ac_num}, are also redisplayed.  To make
connections with the previous analysis regarding convergence, the
ratio of the spring constants for the two states is 1.8 ($k_1 =
45$pN/$\mu$m, $k_2 = 25$pN/$\mu$m) and the probability of lying in
state 1 is 30\%.  These conditions nearly satisfy those displayed in
\fref{fig:deltaA2}~(a) and close to the limit we imposed for
applicability.  From \fref{fig:deltaA2}~(a) we expect the convergence
of both series to be within 1\% of the exact solution.  The analytic
solutions truncated at third order are also displayed in the figure.
The series expressed around the fast-switching limit,
\eref{eqn:thirdorder}, is represented by the red dashed-line and the
series expressed around the slow-switching limit,
\eref{eqn:thirdorder0}, is represented by the blue dashed-line.  As
expected, both series truncated at third order accurately portrays the
numerical solution for case~D.

\begin{figure}
\begin{center}
\centerline{\epsfxsize=3.0truein \epsfbox{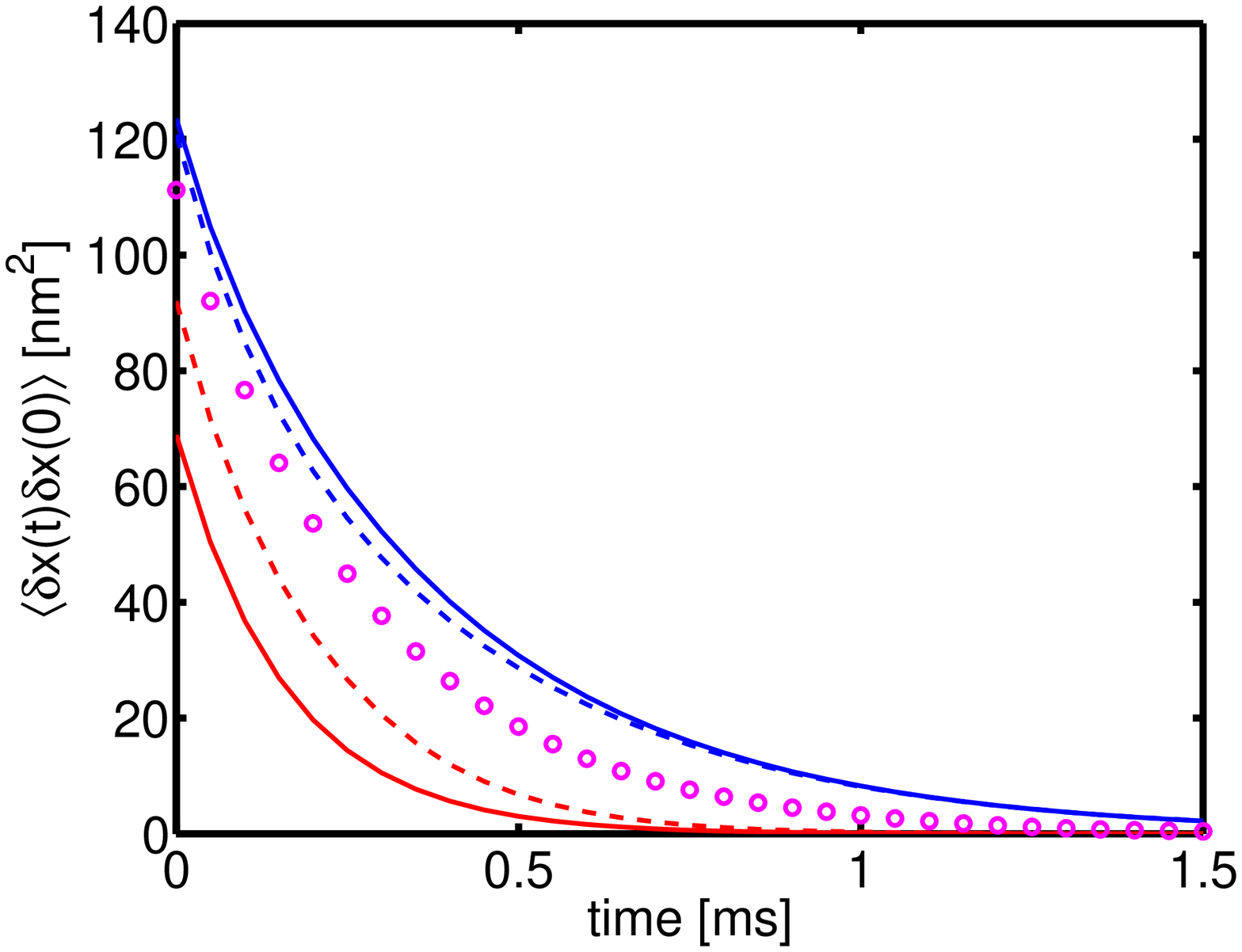}} 
\caption{Auto-correlation function for a system with a fluctuating
spring constant.  Numerical solutions for case~$\mathcal{D}$ is
represented by the red circles. The solid red line is the solution in
the fast-switching limit and the blue line is the solution in the slow
switching limit.  Series solutions for the auto-correlation function,
truncated at third order, are represented by the red dashed-line when
initiated from the fast-switching limit solution
(\eref{eqn:thirdorder}) and by the blue dashed-line when
initiated from the slow-switching limit solution
(\eref{eqn:thirdorder0}).}\label{fig:ac_pert_largek1}
\end{center}
\end{figure}

The molecular parameters expressed in case~D are near the region of
minimal acceptable spatial resolution.  Spatial resolution
was enhanced when the spring constant associated with state 1 was
increased to $k_1 = 140$pN/$\mu$m, which we termed as
case~$\mathcal{D}$.  This results in a spring constant ratio of
$k_1/k_2 = 5.6$, which is somewhat outside the predicted bounds for
applicability.  To test this idea, \fref{fig:ac_pert_largek1} presents
the numerical solutions for case~$\mathcal{D}$ along with the
truncated series solutions.  The numerical solutions are again
represented by red circles, the solutions in the fast-switching limit
is represented by the solid blue line and slow-switching limit is
represented by the solid red line.  The analytic solution expressed
around the fast-switching limit, \eref{eqn:thirdorder}, is represented
by the red dashed-line and the analytic solution expressed around the
slow-switching limit, \eref{eqn:thirdorder0}, is represented by the
blue dashed-line.  As expected, both series truncated at third order
do not accurately portrays the numerical solution for
case~$\mathcal{D}$.

\section{Conclusion} \label{sec:conclusion}

In this work we presented an analytic method applicable to the
analysis of data from single-molecule switching experiments.  This
method regards the analysis of the auto-correlation.  The
auto-correlation function was taken as the vehicle for analysis,
because it varies on time scales governed by the intrinsic time
constant of the experimental probe and; therefore, its functional form
should be effected by biomolecular transitions which occur on such
corresponding time scales.  This is an important attribute, because we
presented arguments (theoretically motivated) to why this is also a
temporal regime inaccessible for exploration by traditional averaging
schemes; expanding the temporal capabilities of single molecule
studies.

To bolster this position, we also presented mock (numerical
simulations) experiments, demonstrating the failure of traditional
averaging methods to predict both the mechanical and kinetic
properties of biomolecular systems, when the transitions between the
biomolecular states occur on the time scales governing the motion of
the experimental probe.  To further motivate the attributes of the
auto-correlation function, we also presented numerical solutions which
demonstrated how its functional form alters in this temporal regime.
To comprehend how the auto-correlation function may be influenced by
the rates of transition, we then presented two physically motivated
guesses for its solution.  We argued that these two guesses should
correspond to solutions for the auto-correlation function in two
physically opposing extremes: when the switching between biomolecular
states is either much faster or much slower than the time scales
governing the motion of the probe.

Motivated by these numerical simulations we then presented our
mathematical analysis for the solution to the auto-correlation
function.  We modeled the experimental probe as an over-damped
particle, in which the associated spring constants corresponding to
the biomolecule fluctuates between different states.  We presented a
formal solution for the auto-correlation function, from which we
proved that the two physically motivated solutions also correspond to
mathematical bounds (opposing) to the general solution.

Although no known ``closed'' form solution exists for the problem at
hand, we presented series solutions for the auto-correlation function.
Inspired by the bounds found for the general solution, two series
solutions for the auto-correlation function were expressed around
these bounds; providing series solutions which converge to the exact
solutions from opposing positions.  In addition, we proved
mathematically that the two bounds indeed correspond to exact
solutions for the auto-correlation function in the anticipated
regimes.

Having obtained two series solutions for the auto-correlation function
we next presented a general analysis regarding their convergence to
the general solution and their applicability to the analysis of
single-molecule switching experiments.  To garner concrete
understanding regarding the convergence and applicability of this
method, we then focuses on the two state problem.  A range of
biomolecular parameters, for where we believe this method would be
applicable for the analysis of experimental data, was then determined.
Guided by this analysis, we then presented solutions for each series
(up to third order) for a two state Markov process.  We compared our
truncated series solutions with numerical solutions for the
auto-correlation function and showed that these truncated series
solutions converged to the numerical solution to the degree
anticipated from our general analysis regarding convergence.  These
truncated series solutions for the auto-correlation function are
amiable to the analysis of experimental data, in the regime where
traditional averaging procedures fail.

The general analysis presented here should be applicable to
single-molecule experimental studies.  Although the range of
parameters found for applicability of this method appears to be
limited, this parameter range is beyond the reach for analysis with
current traditional methods, providing an expanded region of phase
space that can be explored by single-molecule methods.

\begin{acknowledgments}
The author is grateful to R. Phillips and M. Inamdar for useful
discussions.  This work was supported under AFOSR/DARPA grant
F49620-02-1-0085.
\end{acknowledgments}

\appendix

\section{Variance of A Time Averaged Property} \label{app:altarg}

We now present an argument determining the reduction of the width of
the distribution function for time averaged values $\langle
x\rangle_{\rm T}$.  The widths of the distribution functions are
related to the square root of the variance, $\sigma^2$.  Consider
measurements of a property $x$ over a given time window T and the
average value of this measurement, $\langle x\rangle_{\rm T}$, is
tabulated.  The relationship between the data set corresponding to the
time averaged value to that of the data set corresponding to the
instantaneous measurements is characterized by the following equality
\begin{equation}
\left(\begin{array}{c} \av{x}_1 \\ \ \\ \av{x}_2 \\ \vdots \\ \av{x}_{\rm
N}\end{array}\right) = \frac{1}{\text M} \left(\begin{array}{c}
\sum_{j=1}^{\rm M} x_{1,j} \\ \ \\ \sum_{j=1}^{\rm M} x_{2,j} \\ \vdots \\
\sum_{j=1}^{\rm M} x_{N,j} \end{array} \right).
\end{equation}
Here $x_{i,j}$ corresponds to an instantaneous measurement of $x$ and
the time averaged value $\langle x\rangle_i$ corresponds to the
average of $x$ over time window $i$.  Each time window corresponds to
M data points and there are a total of N$\times$M instantaneous
measurements.

Now we make a comparison between the variance corresponding to the
instantaneous measurements (whole data set),
\begin{equation}
\sigma_x^2 = \frac{1}{\rm NM}\sum_{i,j}^{\rm N,M}
\bigl(x_{i,j}-\bar x\bigr)^2,\label{eqn:sig}
\end{equation}
with that of the time averaged value,
\begin{equation}
\sigma_{\langle x\rangle_{\rm T}}^2 = \frac{1}{\rm N}\sum_i^{\rm N}
\left(\av{x}_i - \bar x \right)^2.\label{eqn:sigave}
\end{equation}
Here $\bar x$ is the average value of $x$,
\begin{equation}
\bar x = \frac{1}{NM}\sum_{i,j}^{\rm N,M} x_{i,j}.
\end{equation}

A few assumptions are now made.  First it is assumed that the
instantaneous measurements are uncorrelated, statistically
independent.  This implies that the instantaneous measurements are
stored at time intervals on order of the intrinsic time constant of
the experimental probe, as this is roughly the time scale to make two
uncorrelated measurements.  Next, we assume that the size of the data
set corresponding to the time averaged value (N) is ``large'' such
this data set ($\{\av{x}_i\}$) adequately samples the distribution
space of the time averaged value.  Furthermore, we assume that the
number of data points used to compute each averaged value (M) is
``small'' such that the distribution function corresponding to the
time averaged value contains a measurable width, $\sigma_{\langle
x\rangle_{\rm T}} \neq 0$.  (This also implies N$\gg$M.) Finally, we
assume that all distribution functions have mirror symmetry around the
mean ($\bar x$) of the measured value $x$.  These conditions are
consistent with many single molecule experiments.

Given these assumptions we are now in position to relate the variance
corresponding to the instantaneous measurements, $\sigma^2_x$, to that
corresponding to the time averaged measurements, $\sigma^2_{\langle
x\rangle_{\rm T}}$.  The variance corresponding to the time averaged
value, as expressed in \eref{eqn:sigave}, can be related to the
instantaneous measurements as follows,
\begin{eqnarray}
\sigma_{\langle x\rangle_{\rm T}}^2 &=& \frac{1}{\rm N}\sum_i^{\rm N}
\left(\av{x}_i - \bar x \right)^2 \\ &=& \frac{1}{\rm N} \sum_i^N
\left(\frac{1}{\rm M}\sum_j^{\rm M} x_{i,j} - \bar x \right)^2 \\ &=&
\frac{1}{\rm NM^2}\sum_i^{\rm N} \left(\sum_j^{\rm M}\bigl(x_{i,j} -
\bar x\bigr) \right)^2.\label{eqn:var_a}
\end{eqnarray}
Because of the assumption that M is small, implying that $\sum_j^{\rm
M} x_{i,j}/M \neq \bar x$, the last term does not vanish.  This is
consistent with having a nonzero variance corresponding to the time
averaged value, $\sigma_{\langle x\rangle_{\rm T}} \neq 0$.

Further decomposing, the variance (\eref{eqn:var_a}) can be
written as
\begin{eqnarray}
\sigma_{\langle x\rangle_{\rm T}}^2 &=& \frac{1}{\rm NM^2}\sum_i^{\rm
 N}\sum_j^{\rm M} \bigl(x_{i,j} -\bar x \bigr)^2 \\ && + \frac{1}{\rm
 NM^2} \sum_i^{\rm N}\sum_{j\neq k}^{M,M} \bigl(x_{i,j}-\bar x
 \bigr)\bigl(x_{i,k}-\bar x \bigr).\label{eqn:var_b}
\end{eqnarray}
Next, we use the assumption that N is ``large''. In the limit that
N goes to infinity the last term in \eref{eqn:var_b} vanishes,
\begin{equation}
\lim_{N\rightarrow \infty} \sum_i^{\rm N} \bigl(x_{i,j}-\bar x
 \bigr)\bigl(x_{i,k}-\bar x \bigr) = 0,
\end{equation}
for $j\neq k$. This equality holds when the data set $\av{x}_{\rm T}$
contain a number of statistically independent points and the
distribution function has mirror symmetry around the average value
$\bar x$.

The variance corresponding to the time averaged value now can be
written as
\begin{eqnarray}
\sigma_{\langle x\rangle_{\rm T}}^2 &=& \frac{1}{\rm NM^2}\sum_i^{\rm
 N}\sum_j^{\rm M} \bigl(x_{i,j} -\bar x \bigr)^2 \\ &=& \frac{1}{\rm M}
 \left(\frac{1}{\rm NM} \sum_{i,j}^{\rm N,M} \bigl(x_{i,j} - \bar x
 \bigr)^2\right)\\ &=& \frac{1}{\rm M} \sigma^2_x.
\end{eqnarray}
Revealing that the variance corresponding to the time average value is
equal to the variance corresponding to the instantaneous measurements,
$x$, reduced by the number of uncorrelated measurements used to
determine the time averaged value.

\section{Time constant}\label{app:tc}

The time scales governing the motion of the probe correspond to its
time constant $\tau = \gamma/k$, where $k$ is the spring constant
associated with the end of the probe.  This time scale is also the
time scale at which two uncorrelated measurements are made.  This
spring constant contains contributions from both the intrinsic time
constant of the experimental probe, $k_{\rm P}$, and that from the
molecule in state $i$, $k_{{\rm m}i}$.  This time constant is less
than the intrinsic time constant of the experimental probe, $\tau_{\rm
P} = \gamma/k_{\rm P}$.  For systems of interest the spring constant
of the probe is generally much larger than or of order of the molecule
in any state and; therefore, the intrinsic time constant of the
experimental probe serves as an appropriate measure for the relaxation
time of the probe when attached to a biomolecule, $\tau_{\rm P}
\approx \tau$.  Using the intrinsic time constant of the probe as the
measure of its relaxation also proves to be a convenient measure, as
the particular state of the molecule does not have to be continually
referred to and quantified during this analysis.  (Certainly, the
discussions here can be generalized to the cases when the spring
constant of the molecule in any state is much larger than the intrinsic
spring constant of the probe.)

\section{Auto-correlation function, when $\langle f(t_1)f(t_2)\cdots f(t_n)\rangle$ is independent of time} \label{app:avC}

If all of the moments of the random variable $f$ are independent of
the time variables then \eref{eqn:xx2} takes the form
\begin{eqnarray}
&&\av{\delta x(t)\delta x(0)} = \label{eqn:acf} \\ &&
\frac{\kT}{k_a}\sum_{n,n'=0}^\infty\frac{ C_{n+n'}}{\tau_d^{n+n'}}
\nonumber \int_0^\infty \prod_{i=0}^n dt_i \prod_{i'=0}^{n'} dt_{i'}
e^{-\frac{|t+T_n-T_{n'}|+T_n+T_{n'}}{\tau_a}},
\end{eqnarray}
where
\begin{equation}
C_{n+n'} = \av{\mathcal{F}_n(0,\{T_j\})\mathcal{F}_{n'}(t,\{T_{j'}\})}\nonumber
\end{equation}
is independent of the time variables. It, therefore, only depends on
the sum of $n$ and $n'$ and not their individual values.

To evaluate this integral we first calculate the Fourier transform of
the auto-correlation function and then integrate over the time
variables. We will then take the inverse transform by using the
residue theorem for complex integration. Finally, through algebraic
identities we obtain the result in \eref{eqn:xxC}.

The Fourier transform $\hat f(\omega)$ of a function $f(t)$ is defined
as
\begin{eqnarray}
\hat f(\omega) &=& \int_{-\infty}^\infty dt f(t) e^{-i\omega t}\nonumber
\end{eqnarray}
and the inverse transform is defined as
\begin{eqnarray}
f(t) &=& \frac{1}{2\pi} \int_{-\infty}^\infty d\omega \hat f(\omega)
e^{i\omega t}.\nonumber
\end{eqnarray}
The Fourier transform of the auto-correlation function in this
limiting form  is
\begin{eqnarray}
\av{\hat x(\omega)x(0)} &=&
\frac{\kT}{k_a}\sum_{n,n'=0}^\infty\left(\frac{1}{\tau_d^{n+n'}}\right)
  \frac{2C_{n+n'}/\tau_a}{(1/\tau_a)^2 + \omega^2}
\nonumber \\ && \nonumber \times\int_0^\infty \prod_{i=0}^n dt_i
e^{-t_i(\frac{1}{\tau_a} - i\omega)} 
\prod_{i'=0}^{n'} dt_{i'}e^{-t_i'(\frac{1}{\tau_a} + i\omega)}.
\end{eqnarray}
Integrating over all the time variables, $\{t_i\}$ and $\{t_{i'}\}$, the
Fourier transform becomes
\begin{eqnarray}
\av{\hat x(\omega)x(0)} &=&\label{eqn:ftxx}
\frac{\kT}{k_a}\sum_{n,n'=0}^\infty\left(\frac{1}{\tau_d^{n+n'}}\right)
\frac{C_{n+n'}}{\tau_a/2} \\ && \nonumber \times\left(\frac{i}{\omega
+ i/\tau_a} \right)^{n+1} \left(\frac{-i}{\omega - i/\tau_a}
\right)^{n'+1}.
\end{eqnarray}
Taking the inverse transform of \eref{eqn:ftxx} the
auto-correlation function is
\begin{eqnarray}
\av{\delta x(t)\delta x(0)} &=&
\frac{\kT}{k_a}\sum_{n,n'=0}^\infty\left(\frac{1}{\tau_d^{n+n'}}
\right) \frac{C_{n+n'}}{\tau_a/2} \int_{-\infty}^\infty d\omega
e^{i\omega t} \nonumber \\ &&
\times\frac{1}{2\pi}\left(\frac{i}{\omega + i/\tau_a} \right)^{n+1}
\left(\frac{-i}{\omega - i/\tau_a} \right)^{n'+1} \nonumber
\end{eqnarray}
or
\begin{eqnarray}
\av{\delta x(t)\delta x(0)}& =
&\frac{\kT}{k_a}\sum_{n,n'=0}^\infty\left(\frac{i}{\tau_d}
\right)^{n+n'} \frac{C_{n+n'}}{\tau_a\pi}
\left(\frac{\tau_a}{i^3}\right)^{n'}\frac{1}{n'!} \nonumber \\ &&
\left.  \times\frac{\partial^{n'}}{\partial a^{n'}}
\int_{-\infty}^\infty d\omega \label{eqn:inter}
e^{i\omega t}\frac{\left(\omega + i/\tau_a\right)^{-n-1}}{\omega - ia/\tau_a}
\right|_{a=1}. 
\end{eqnarray}

Equation~\ref{eqn:inter} can be integrate in the upper-half
complex-plane. The integrand has a pole at $\omega = ia/\tau_a$ and
\begin{eqnarray}
\av{\delta x(t)\delta x(0)} &=&
\frac{\kT}{k_a}\sum_{n,n'=0}^\infty 2\left(\frac{\tau_a}{\tau_d}
\right)^{n+n'} C_{n+n'} \frac{(-1)^{n'}}{n'!}\nonumber \\&&
\left. \times\frac{\partial^{n'}}{\partial a^{n'}}
\frac{e^{-at/\tau_a}}{(1+a)^{n+1}} \right|_{a=1}. \label{eqn:ftxxw}
\end{eqnarray}
Using the identity
\begin{equation}
 \frac{\partial^{n'}}{\partial a^{n'}} AB = \sum_{j=0}^{n'}
 \frac{n'!}{j!(n'-j)!} A^{[j]}B^{[n'-j]},\nonumber
\end{equation}
where $X^{[j]}$ is the $j^{th}$ derivative of $X$, and switching
variables to $m=n+n'$ and $n'$, \eref{eqn:ftxxw} can be
written as
\begin{eqnarray}
\av{\delta x(t)\delta x(0)} &=& 
\frac{\kT}{k_a} e^{-t/\tau_a}\sum_{m=0}^\infty
\left(\frac{\tau_a}{\tau_d}\right)^m C_m \label{eqn:foo} \\ &&
\times\sum_{n'=0}^m \sum_{j=0}^{n'}
\frac{(m-j)!}{j!(n'-j)!(m-n')!}\left(\frac{t}{\tau_a}\right)^j
\frac{1}{2^{m-j}}.  \nonumber
\end{eqnarray} 
Using the identity
\begin{equation}
\sum_{n'=0}^m \sum_{j=0}^{n'} = \sum_{j=0}^m\sum_{n'=j}^m\nonumber
\end{equation}
Equation~\ref{eqn:foo} can be written as
\begin{widetext}
\begin{eqnarray}
\av{\delta x(t)\delta x(0)} &=& \frac{\kT}{k_a}
e^{-t/\tau_a}\sum_{m=0}^\infty \left(\frac{\tau_a}{\tau_d}\right)^m
C_m \sum_{j=0}^m \left(\frac{t}{\tau_a}\right)^j 
\frac{1}{2^{m-j}}
\frac{(m-j)!}{j!}\sum_{n'=j}^{m}\frac{1}{(n'-j)!(m-n')!}  \nonumber \\
&=& \frac{\kT}{k_a} e^{-t/\tau_a}\sum_{m=0}^\infty
\left(\frac{\tau_a}{\tau_d}\right)^m C_m \sum_{j=0}^m
\left(\frac{t}{\tau_a}\right)^j 
\frac{1}{2^{m-j}}
\frac{(m-j)!}{j!}\sum_{l=0}^{m-j}\frac{1}{(m-j-l)!l!}  \nonumber \\
&=& \frac{\kT}{k_a} e^{-t/\tau_a}\sum_{m=0}^\infty
\left(\frac{\tau_a}{\tau_d}\right)^m C_m \sum_{j=0}^m
\frac{1}{j!}\left(\frac{t}{\tau_a}\right)^j,
\end{eqnarray}
\end{widetext}
the desired result.

\section{Expansion for a two state Markov process} \label{app:markov}

We present solutions for the series expansions represented in
\erefs{eqn:xx2inf}~and~\ref{eqn:xx20}, up to third order, for a system
that behaves as a two state Markov process.  Here, a biomolecule is
considered a two state Markov process if two distinct mechanical
states exist (different spring constants) and the probability of
making a transition from one state to another is local in time
(independent of all previous configurations).  We provide closed form
solutions for the auto-correlation function (truncated at third order)
that depends on the magnitude of the rates and can be used in the
analysis of experimental data.  We proceed by first reviewing the
relevant theory of a two state Markov process.

\subsection{The two state problem: Statistics} \label{app:twostate}

We review some statistics governing a two state Markov process.
Particularly, properties required for the evaluation of the moments
($\Di{n,n'}, \Dz{n,n'}$) expressed in the solution of the
auto-correlation function,
\erefs{eqn:xx2inf}~and~\ref{eqn:xx20}.

Consider a two state Markov process and assume that at time $t = 0$
the system starts in state $i$~\footnote{$i$ can be either state 1 or
2}. The probability of the system lying in state $j$ at time t obeys
the following rate equation
\begin{equation}
\frac{\partial}{\partial t}\left(\begin{array}{l}P_{i1}(t) \\ P_{i2}(t)
\end{array} \right) = \left(\begin{array}{rr}-\kappa_{12} & \kappa_{21}\\
\kappa_{12} & -\kappa_{21} \end{array} \right)\left(\begin{array}{l}P_{i1}(t)
\\ P_{i2}(t) \end{array} \right).\label{eqn:Pd}
\end{equation}
Here $P_{ij}(t)$ is the probability of the system lying in state $j$
at time $t$ if at time $t = 0$ the system was in state $i$, $\kappa_{21}$ is
the rate of going into state 1 from state 2 and $\kappa_{12}$ is the rate
of going into state 2 from state 1.

Define the rate matrix as
\begin{equation*}
\mathbf{W} \equiv \left(\begin{array}{rr}-\kappa_{12} & \kappa_{21}\\
  \kappa_{12} & -\kappa_{21} \end{array} \right).
\end{equation*}
The probability of the system lying in a particular state ($P_{ij}$)
at time $t$ follows from solving the differential equation represented
in \eref{eqn:Pd}. In matrix notation the solution is
\begin{equation}
\vec{P}(t) = \mathbf{T}_t \vec{P}(0), \label{eqn:Pt}
\end{equation}
Here the transition probability matrix, $\mathbf{T}_t$, is defined as
\begin{equation}
\mathbf{T}_t \equiv e^{\mathbf{W}t},
\end{equation}
whose ($i,j$) component, $T_{t,ij}$, is defined as the probability of
the system lying in state $i$ at time $t$ if it was in state $j$ at
time 0.  Because we assume a Markov process, the stationary solution
to \eref{eqn:Pt} and the transition probability matrix is all
that is needed to obtain the statistics governing the transitions of
the system.  All higher order, multi-variable, joint probability
distributions functions can be obtained from them.

The stationary solution to \eref{eqn:Pt} is
\begin{equation}
\vec{P} = \left(\begin{array}{c} P_1 \\ P_2\end{array}\right),
\end{equation}
where
\begin{eqnarray*}
P_1 &=& \kappa_{21}/\kapt, \\ P_2 &=& \kappa_{12}/\kapt
\end{eqnarray*}
and
\begin{equation}
\kapt = \kappa_{21} + \kappa_{12}.
\end{equation}
It is straightforward to show that the transition probability matrix
is
\begin{equation*}
\mathbf{T}_t = \mathbf{1} - (1-e^{-\kapt t})\left(\begin{array}{rr}  P_2 & -P_1 \\ -P_2 &   P_1
\end{array}\right),
\end{equation*}
where $\mathbf{1}$ is the identity matrix.

For the moment we refine our discussion to instances of time.  The
stochastic variable of interest is the fluctuating term $f(t)$,
\eref{eqn:fi}.  At time $t_n$ the value of the fluctuating term is
denoted as $f^{(n)}$, which can take one of two values: $f_1$ if in
state 1 and $f_2$ if in state 2.  For the problem at hand these values
are $f_i = \mp 1$, where the minus sign corresponds to the system with
the larger spring constant, (see \eref{eqn:fi}).  Because we now
consider instances of time we define slight variants to the stationary
probability solution, $\vec{P}$, and transition probability matrix,
$\mathbf{T}_t$.  These variants refer to particular values of $f$
instead of particular states of the system.  The stationary
probability of $f$ having value $f^{(n)}$ is denoted as
$\Bbb{P}_1(f^{(n)})$ (single variable probability distribution
function) .  The probability of $f$ having value $f^{(\alpha)}$ at
time $t_\alpha$ if it have had value $f^{(\beta)}$ at time $t_\beta$
is denoted as $T_{t_{\alpha,\beta}}(f^{(\alpha)}|f^{(\beta)})$.

The multi-variable joint probability distribution functions, defined
as the probability of $f$ having value $f^{(n)}$ at time $t_n$ if $f$
had value $f^{(n-1)}$ at time $t_{n-1}$ and value $f^{(n-2)}$ at time
$t_{n-2}$ and so on, is
\begin{eqnarray*}
&&\Bbb{P}_n(f^{(1)},t_1;f^{(2)},t_2;\cdots;f^{(n)},t_n) =\\&&
T_{t_{n,n-1}}(f^{(n)}|f^{(n-1)})\cdots T_{t_{3,2}}(f^{(3)}|f^{(2)})T_{t_{2,1}}(f^{(2)}|f^{(1)})\\&&\times\Bbb{P}_1(f^{(1)}).
\end{eqnarray*}
Here time is ordered, $t_1 \leq t_2 \leq ....\leq t_n$, and $t_{m,m-1}
\equiv t_m-t_{m-1} \geq 0$ is the time difference of neighboring time
instances.

Averages over the stochastic variable $f$ are needed to evaluate the
integrals in \erefs{eqn:xx2inf}~and~\ref{eqn:xx20}.  They can be
written as
\begin{eqnarray*}
&&\av{f(t_1)f(t_2)...f(t_n)} = \\ &&
\sum_{f^{(1)}\cdots f^{(n)}=\{f_1,f_2\}} f^{(1)} f^{(2)} \cdots f^{(n)}
\Bbb{P}_n(f^{(1)},t_1;\cdots;f^{(n)},t_n)
\end{eqnarray*}
or in an alternative matrix form
\begin{equation}
\av{f(t_1)f(t_2)...f(t_n)} = \bigl(f_1 \ \
f_2\bigr)\biggl(\prod_{i=1}^{n-1}\bigl(\mathbf{T}_1 +
e^{-\kapt t_{i,i-1}}\mathbf{T}_2\bigr)\biggr)\vec{P}, \label{eqn:avf}
\end{equation}
where $t_1 \leq t_2 \leq ...\leq t_n$ (time ordering).  Here
\begin{eqnarray}
\mathbf{T}_1 &=& \left(\begin{array}{cc}f_1P_1 & f_2P_1 \\ f_1P_2 & f_2P_2
\end{array}\right), \label{eqn:foo1}\\
\mathbf{T}_2 &=& \left(\begin{array}{cc}f_1P_2 & -f_2P_1 \\ -f_1P_2 &
f_2P_1
\end{array}\right).\label{eqn:foo2} 
\end{eqnarray}
Given the averages of the fluctuating term, \eref{eqn:avf}, terms
in the series expansion can be evaluated for the general case.

\subsection{Truncated series}
Using the definitions above, we calculated the general solution for the
auto-correlation, up to third order.  Expanding around the
fast-switching limit the auto-correlation function takes the form
\begin{widetext}
\begin{equation}
\av{x(t)x(0)}_3^{(\infty)} = \frac{\kT}{\av{k}}e^{-t\langle
\tau^{-1}\rangle} +
\frac{\kT}{k_a}\left(\frac{\tau_a}{\tau_d}\right)^24P_1P_2e^{-t/\tau_a}
\left[ \left(\vec C_2 + e^{-\kapt
t}\vec D_2\right)^\dagger    \vec T_2(t) 
+\frac{\tau_a}{\tau_d} \av{f}\left(\vec C_3 + e^{-\kapt t}\vec D_3\right)^\dagger    \vec
T_3(t)\right], \label{eqn:thirdorder}
\end{equation}
\end{widetext}
where the superscript ($(\infty)$) denotes that the series is expanded
around the fast-switching limit and the subscript denotes the order at
which the series is truncated.  In \eref{eqn:thirdorder} we present
new notation.  The vector $\vec T_n(t)$ is a $n$-component vector that
is a function of time and defined as
\begin{equation}
\vec T_n(t) = \left(\begin{array}{c}1\\ \frac{t}{\tau_a} \\ \vdots \\
\frac{1}{(n-1)!}\left(\frac{t}{\tau_a}\right)^{(n-1)}\end{array} \right).
\end{equation}
The coefficient vectors, $\vec C_n$ and $\vec D_n$, are functions of
$\kapt\tau_a$, whose values are listed in Table~\ref{tab:coef}.
Dagger ($\dagger$) represents the transpose of a vector and $\av{f} =
f_1P_1+f_2P_2$ is the average value of the fluctuating variable.  The
subscripts on the vectors $\vec T_n(t), \ \vec C_n$ and $\vec D_n$
represents the order each term was calculated, second or third, and
also corresponds to the size of the vector.  In limit that $\kttau
\rightarrow \infty$ the coefficient vectors in
\eref{eqn:thirdorder} vanish, leaving only the fast-switching
limit solution.

As expressed in \sref{sec:convergence}, once the fast-switching
limit is determined expansion around the slow-switching limit is
easily determined by exploiting
\erefs{eqn:thirdorder}~and~\ref{eqn:Amt}:
\begin{widetext}
\begin{equation}
\av{x(t)x(0)}_3^{(0)} =
\left\langle\frac{\kT}{k}e^{-t/\tau}\right\rangle +
\frac{\kT}{k_a}\left(\frac{\tau_a}{\tau_d}\right)^24P_1P_2e^{-t/\tau_a}\left[
\left(\vec C_2 + e^{-\kapt t}\vec{D}_2 - 1\right)^\dagger \vec T_2(t)
+\frac{\tau_a}{\tau_d} \av{f}\left(\vec C_3 + e^{-\kapt
t}\vec{D}_3 - 1\right)^\dagger \vec T_3(t) \right],
\label{eqn:thirdorder0}
\end{equation} 
\end{widetext}
where the superscript ($(0)$) denotes that the series is expanded
around the slow-switching limit.  Here, in the limit that $\kttau
\rightarrow 0$ the terms in the square brackets sum to zero, leaving
the slow-switching limit solution.

\begin{table}
\caption{Coefficients for two state Markov process in the
fast-switching limit.  The indices for the vectors $\vec C$ and $
\vec{D}$ are denoted by the subscripts.  The coefficients
for the slow-switching limit is determined trivial through these
coefficients and the use of \eref{eqn:Amt}. \label{tab:coef}}
\begin{ruledtabular}
\begin{tabular}{cc} 
Coefficients & Value ($\alpha=\kttau$) \vspace{.1cm}\\ \hline
$C_{2,1}$ & $\frac{2\alpha^2 + \alpha - 2
}{(2+\alpha)\alpha^2}$ \vspace{.1cm} \\ $C_{2,2}$  &
$\frac{1}{\alpha}$  \vspace{.1cm}\\ 
$D_{2,1}$  & $\frac{2-\alpha}{(2+\alpha)\alpha^2}$ \vspace{.1cm}\\
$D_{2,2}$  & 0 \vspace{.1cm}\\
 $C_{3,1}$  & $\frac{(17/4)\alpha^4 + (11/2)\alpha^3 -
4\alpha^2 + 4\alpha + 16}{(2+\alpha)^2\alpha^3}$  \vspace{.1cm}\\ 
$C_{3,2}$  & $\frac{(7/2)\alpha^2 - 6}{(2+\alpha)\alpha^2}$ \vspace{.1cm}\\
 $C_{3,3}$  & $\frac{2}{\alpha}$ 
\vspace{.1cm}\\  $D_{3,1}$  & $\frac{-\alpha^3 + 4\alpha^2 -
4\alpha - 16}{(2+\alpha)^2\alpha^3}$ \vspace{.1cm}\\  $D_{3,2}$
 & $\frac{\alpha - 2}{(2+\alpha)\alpha^2}$\vspace{.1cm}\\
$D_{3,3}$  & 0 
\end{tabular}
\end{ruledtabular}
\end{table}

Equations~\ref{eqn:thirdorder}~and~\ref{eqn:thirdorder0} are general
solutions for the auto-correlation function.  These solutions depend
on the magnitude of the rates of transition for the molecule,
$\kappa_{12}$ and $\kappa_{21}$, unlike the solutions in the
fast-switching and slow-switching limits, which depend only on the
ratio of the rates. 

\bibliographystyle{apsrev} \bibliography{sinmolexp_protfold}

\begin{thebibliography}{36}
\expandafter\ifx\csname natexlab\endcsname\relax\def\natexlab#1{#1}\fi
\expandafter\ifx\csname bibnamefont\endcsname\relax
  \def\bibnamefont#1{#1}\fi
\expandafter\ifx\csname bibfnamefont\endcsname\relax
  \def\bibfnamefont#1{#1}\fi
\expandafter\ifx\csname citenamefont\endcsname\relax
  \def\citenamefont#1{#1}\fi
\expandafter\ifx\csname url\endcsname\relax
  \def\url#1{\texttt{#1}}\fi
\expandafter\ifx\csname urlprefix\endcsname\relax\def\urlprefix{URL }\fi
\providecommand{\bibinfo}[2]{#2}
\providecommand{\eprint}[2][]{\url{#2}}

\bibitem[{\citenamefont{Finzi and Gelles}(1995)}]{finzi95}
\bibinfo{author}{\bibfnamefont{L.}~\bibnamefont{Finzi}} \bibnamefont{and}
  \bibinfo{author}{\bibfnamefont{J.}~\bibnamefont{Gelles}},
  \bibinfo{journal}{Science} \textbf{\bibinfo{volume}{267}},
  \bibinfo{pages}{378} (\bibinfo{year}{1995}).

\bibitem[{\citenamefont{Smith et~al.}(1996)\citenamefont{Smith, Cui, and
  Bustamante}}]{smith96}
\bibinfo{author}{\bibfnamefont{S.~B.} \bibnamefont{Smith}},
  \bibinfo{author}{\bibfnamefont{Y.}~\bibnamefont{Cui}}, \bibnamefont{and}
  \bibinfo{author}{\bibfnamefont{C.}~\bibnamefont{Bustamante}},
  \bibinfo{journal}{Science} \textbf{\bibinfo{volume}{271}},
  \bibinfo{pages}{795} (\bibinfo{year}{1996}).

\bibitem[{\citenamefont{Mehta et~al.}(1997)\citenamefont{Mehta, Finer, and
  Spudich}}]{mehta97}
\bibinfo{author}{\bibfnamefont{A.~D.} \bibnamefont{Mehta}},
  \bibinfo{author}{\bibfnamefont{J.~T.} \bibnamefont{Finer}}, \bibnamefont{and}
  \bibinfo{author}{\bibfnamefont{J.~A.} \bibnamefont{Spudich}},
  \bibinfo{journal}{Proc. Natl. Acad. Sci. USA} \textbf{\bibinfo{volume}{94}},
  \bibinfo{pages}{7927} (\bibinfo{year}{1997}).

\bibitem[{\citenamefont{Rief et~al.}(1997{\natexlab{a}})\citenamefont{Rief,
  Oesterhelt, Heymann, and Gaub}}]{rief97a}
\bibinfo{author}{\bibfnamefont{M.}~\bibnamefont{Rief}},
  \bibinfo{author}{\bibfnamefont{F.}~\bibnamefont{Oesterhelt}},
  \bibinfo{author}{\bibfnamefont{B.}~\bibnamefont{Heymann}}, \bibnamefont{and}
  \bibinfo{author}{\bibfnamefont{H.~E.} \bibnamefont{Gaub}},
  \bibinfo{journal}{Science} \textbf{\bibinfo{volume}{275}},
  \bibinfo{pages}{1295} (\bibinfo{year}{1997}{\natexlab{a}}).

\bibitem[{\citenamefont{Rief et~al.}(1997{\natexlab{b}})\citenamefont{Rief,
  Gautel, Oesterhelt, Fernandez, and Gaub}}]{rief97b}
\bibinfo{author}{\bibfnamefont{M.}~\bibnamefont{Rief}},
  \bibinfo{author}{\bibfnamefont{M.}~\bibnamefont{Gautel}},
  \bibinfo{author}{\bibfnamefont{F.}~\bibnamefont{Oesterhelt}},
  \bibinfo{author}{\bibfnamefont{J.~M.} \bibnamefont{Fernandez}},
  \bibnamefont{and} \bibinfo{author}{\bibfnamefont{H.~E.} \bibnamefont{Gaub}},
  \bibinfo{journal}{Science} \textbf{\bibinfo{volume}{276}},
  \bibinfo{pages}{1109} (\bibinfo{year}{1997}{\natexlab{b}}).

\bibitem[{\citenamefont{Guilford et~al.}(1997)\citenamefont{Guilford, Dupuis,
  ans J.~Wu, Patlak, and Warshaw}}]{guilford97}
\bibinfo{author}{\bibfnamefont{W.~H.} \bibnamefont{Guilford}},
  \bibinfo{author}{\bibfnamefont{D.~E.} \bibnamefont{Dupuis}},
  \bibinfo{author}{\bibfnamefont{G.~K.} \bibnamefont{ans J.~Wu}},
  \bibinfo{author}{\bibfnamefont{J.~B.} \bibnamefont{Patlak}},
  \bibnamefont{and} \bibinfo{author}{\bibfnamefont{D.~M.}
  \bibnamefont{Warshaw}}, \bibinfo{journal}{Biophys. J.}
  \textbf{\bibinfo{volume}{72}}, \bibinfo{pages}{1006} (\bibinfo{year}{1997}).

\bibitem[{\citenamefont{Oberhauser et~al.}(1998)\citenamefont{Oberhauser,
  Marszalek, Erickson, and Fernandez}}]{oberhauser98}
\bibinfo{author}{\bibfnamefont{A.~F.} \bibnamefont{Oberhauser}},
  \bibinfo{author}{\bibfnamefont{P.~E.} \bibnamefont{Marszalek}},
  \bibinfo{author}{\bibfnamefont{H.~P.} \bibnamefont{Erickson}},
  \bibnamefont{and} \bibinfo{author}{\bibfnamefont{J.~M.}
  \bibnamefont{Fernandez}}, \bibinfo{journal}{Nature}
  \textbf{\bibinfo{volume}{393}}, \bibinfo{pages}{181} (\bibinfo{year}{1998}).

\bibitem[{\citenamefont{Veigel et~al.}(1998)\citenamefont{Veigel, Bartoo,
  White, Sparrow, and Molloy}}]{veigel98}
\bibinfo{author}{\bibfnamefont{C.}~\bibnamefont{Veigel}},
  \bibinfo{author}{\bibfnamefont{M.~L.} \bibnamefont{Bartoo}},
  \bibinfo{author}{\bibfnamefont{D.~C.~S.} \bibnamefont{White}},
  \bibinfo{author}{\bibfnamefont{J.~C.} \bibnamefont{Sparrow}},
  \bibnamefont{and} \bibinfo{author}{\bibfnamefont{J.~E.}
  \bibnamefont{Molloy}}, \bibinfo{journal}{Biophys. J.}
  \textbf{\bibinfo{volume}{75}}, \bibinfo{pages}{1424} (\bibinfo{year}{1998}).

\bibitem[{\citenamefont{Carrion-Vazquez
  et~al.}(1999)\citenamefont{Carrion-Vazquez, Marszalek, Oberhauser, and
  Fernandez}}]{carrionvazquez99}
\bibinfo{author}{\bibfnamefont{M.}~\bibnamefont{Carrion-Vazquez}},
  \bibinfo{author}{\bibfnamefont{P.~E.} \bibnamefont{Marszalek}},
  \bibinfo{author}{\bibfnamefont{A.~F.} \bibnamefont{Oberhauser}},
  \bibnamefont{and} \bibinfo{author}{\bibfnamefont{J.~M.}
  \bibnamefont{Fernandez}}, \bibinfo{journal}{Proc. Natl. Acad. Sci. USA}
  \textbf{\bibinfo{volume}{96}}, \bibinfo{pages}{11288} (\bibinfo{year}{1999}).

\bibitem[{\citenamefont{Marszalek et~al.}(1999)\citenamefont{Marszalek, Lu, Li,
  Carrion-Vazquez, Oberhauser, Schulten, and Fernandez}}]{marszalek99}
\bibinfo{author}{\bibfnamefont{P.~E.} \bibnamefont{Marszalek}},
  \bibinfo{author}{\bibfnamefont{H.}~\bibnamefont{Lu}},
  \bibinfo{author}{\bibfnamefont{H.}~\bibnamefont{Li}},
  \bibinfo{author}{\bibfnamefont{M.}~\bibnamefont{Carrion-Vazquez}},
  \bibinfo{author}{\bibfnamefont{A.~F.} \bibnamefont{Oberhauser}},
  \bibinfo{author}{\bibfnamefont{K.}~\bibnamefont{Schulten}}, \bibnamefont{and}
  \bibinfo{author}{\bibfnamefont{J.~M.} \bibnamefont{Fernandez}},
  \bibinfo{journal}{Nature} \textbf{\bibinfo{volume}{402}},
  \bibinfo{pages}{100} (\bibinfo{year}{1999}).

\bibitem[{\citenamefont{Rief et~al.}(1999)\citenamefont{Rief, Pascual, Saraste,
  and Gaub}}]{rief99}
\bibinfo{author}{\bibfnamefont{M.}~\bibnamefont{Rief}},
  \bibinfo{author}{\bibfnamefont{J.}~\bibnamefont{Pascual}},
  \bibinfo{author}{\bibfnamefont{M.}~\bibnamefont{Saraste}}, \bibnamefont{and}
  \bibinfo{author}{\bibfnamefont{H.~E.} \bibnamefont{Gaub}},
  \bibinfo{journal}{J. Mol. Biol.} \textbf{\bibinfo{volume}{286}},
  \bibinfo{pages}{553} (\bibinfo{year}{1999}).

\bibitem[{\citenamefont{Veigel et~al.}(1999)\citenamefont{Veigel, Coluccio,
  Jontes, Sparrow, Milligan, and Molloy}}]{veigel99}
\bibinfo{author}{\bibfnamefont{C.}~\bibnamefont{Veigel}},
  \bibinfo{author}{\bibfnamefont{L.~M.} \bibnamefont{Coluccio}},
  \bibinfo{author}{\bibfnamefont{J.~D.} \bibnamefont{Jontes}},
  \bibinfo{author}{\bibfnamefont{J.~C.} \bibnamefont{Sparrow}},
  \bibinfo{author}{\bibfnamefont{R.~A.} \bibnamefont{Milligan}},
  \bibnamefont{and} \bibinfo{author}{\bibfnamefont{J.~E.}
  \bibnamefont{Molloy}}, \bibinfo{journal}{Nature}
  \textbf{\bibinfo{volume}{398}}, \bibinfo{pages}{530} (\bibinfo{year}{1999}).

\bibitem[{\citenamefont{Meiners and Quake}(2000)}]{meiners00}
\bibinfo{author}{\bibfnamefont{J.~C.} \bibnamefont{Meiners}} \bibnamefont{and}
  \bibinfo{author}{\bibfnamefont{S.~R.} \bibnamefont{Quake}},
  \bibinfo{journal}{Phys. Rev. Lett} \textbf{\bibinfo{volume}{84}},
  \bibinfo{pages}{5014} (\bibinfo{year}{2000}).

\bibitem[{\citenamefont{Oberhauser et~al.}(2001)\citenamefont{Oberhauser,
  Hansma, Carrion-Vazquez, and Fernandez}}]{oberhauser01}
\bibinfo{author}{\bibfnamefont{A.~F.} \bibnamefont{Oberhauser}},
  \bibinfo{author}{\bibfnamefont{P.~K.} \bibnamefont{Hansma}},
  \bibinfo{author}{\bibfnamefont{M.}~\bibnamefont{Carrion-Vazquez}},
  \bibnamefont{and} \bibinfo{author}{\bibfnamefont{J.~M.}
  \bibnamefont{Fernandez}}, \bibinfo{journal}{Proc. Natl. Acad. Sci. USA}
  \textbf{\bibinfo{volume}{98}}, \bibinfo{pages}{468} (\bibinfo{year}{2001}).

\bibitem[{\citenamefont{Liphardt et~al.}(2001)\citenamefont{Liphardt, Onoa,
  Smith, Jr., and Bustamante}}]{liphardt01}
\bibinfo{author}{\bibfnamefont{J.}~\bibnamefont{Liphardt}},
  \bibinfo{author}{\bibfnamefont{B.}~\bibnamefont{Onoa}},
  \bibinfo{author}{\bibfnamefont{S.~B.} \bibnamefont{Smith}},
  \bibinfo{author}{\bibfnamefont{I.~T.} \bibnamefont{Jr.}}, \bibnamefont{and}
  \bibinfo{author}{\bibfnamefont{C.}~\bibnamefont{Bustamante}},
  \bibinfo{journal}{Science} \textbf{\bibinfo{volume}{292}},
  \bibinfo{pages}{733} (\bibinfo{year}{2001}).

\bibitem[{\citenamefont{Marszalek et~al.}(2002)\citenamefont{Marszalek, Li,
  Oberhauser, and Fernandez}}]{marszalek02}
\bibinfo{author}{\bibfnamefont{P.~E.} \bibnamefont{Marszalek}},
  \bibinfo{author}{\bibfnamefont{H.}~\bibnamefont{Li}},
  \bibinfo{author}{\bibfnamefont{A.~F.} \bibnamefont{Oberhauser}},
  \bibnamefont{and} \bibinfo{author}{\bibfnamefont{J.~M.}
  \bibnamefont{Fernandez}}, \bibinfo{journal}{Proc. Natl. Acad. Sci. USA}
  \textbf{\bibinfo{volume}{99}}, \bibinfo{pages}{4278} (\bibinfo{year}{2002}).

\bibitem[{\citenamefont{Schwaiger et~al.}(2002)\citenamefont{Schwaiger,
  Sattler, Hostetter, and Rief}}]{schwaiger02}
\bibinfo{author}{\bibfnamefont{I.}~\bibnamefont{Schwaiger}},
  \bibinfo{author}{\bibfnamefont{C.}~\bibnamefont{Sattler}},
  \bibinfo{author}{\bibfnamefont{D.~R.} \bibnamefont{Hostetter}},
  \bibnamefont{and} \bibinfo{author}{\bibfnamefont{M.}~\bibnamefont{Rief}},
  \bibinfo{journal}{Nature Materials} \textbf{\bibinfo{volume}{1}},
  \bibinfo{pages}{232} (\bibinfo{year}{2002}).

\bibitem[{\citenamefont{Baker et~al.}(2002)\citenamefont{Baker, Brosseau, Joel,
  and Warshaw}}]{baker02}
\bibinfo{author}{\bibfnamefont{J.~E.} \bibnamefont{Baker}},
  \bibinfo{author}{\bibfnamefont{C.}~\bibnamefont{Brosseau}},
  \bibinfo{author}{\bibfnamefont{P.~E.} \bibnamefont{Joel}}, \bibnamefont{and}
  \bibinfo{author}{\bibfnamefont{D.~M.} \bibnamefont{Warshaw}},
  \bibinfo{journal}{Biophys. J.} \textbf{\bibinfo{volume}{82}},
  \bibinfo{pages}{2134} (\bibinfo{year}{2002}).

\bibitem[{\citenamefont{Lia et~al.}(2003)\citenamefont{Lia, Bensimon,
  Croquette, Allemand, Dunlap, Lewis, Adhya, and Finzi}}]{lia03}
\bibinfo{author}{\bibfnamefont{G.}~\bibnamefont{Lia}},
  \bibinfo{author}{\bibfnamefont{D.}~\bibnamefont{Bensimon}},
  \bibinfo{author}{\bibfnamefont{V.}~\bibnamefont{Croquette}},
  \bibinfo{author}{\bibfnamefont{J.-F.} \bibnamefont{Allemand}},
  \bibinfo{author}{\bibfnamefont{D.}~\bibnamefont{Dunlap}},
  \bibinfo{author}{\bibfnamefont{D.~E.~A.} \bibnamefont{Lewis}},
  \bibinfo{author}{\bibfnamefont{S.}~\bibnamefont{Adhya}}, \bibnamefont{and}
  \bibinfo{author}{\bibfnamefont{L.}~\bibnamefont{Finzi}},
  \bibinfo{journal}{Proc. Natl. Acad. Sci. USA} \textbf{\bibinfo{volume}{100}},
  \bibinfo{pages}{11373} (\bibinfo{year}{2003}).

\bibitem[{\citenamefont{Jacob et~al.}(1999)\citenamefont{Jacob, Holtermann,
  Perl, Reinstein, Schindler, Geeves, and Schmid}}]{jacob99}
\bibinfo{author}{\bibfnamefont{M.}~\bibnamefont{Jacob}},
  \bibinfo{author}{\bibfnamefont{G.}~\bibnamefont{Holtermann}},
  \bibinfo{author}{\bibfnamefont{D.}~\bibnamefont{Perl}},
  \bibinfo{author}{\bibfnamefont{J.}~\bibnamefont{Reinstein}},
  \bibinfo{author}{\bibfnamefont{T.}~\bibnamefont{Schindler}},
  \bibinfo{author}{\bibfnamefont{M.~A.} \bibnamefont{Geeves}},
  \bibnamefont{and} \bibinfo{author}{\bibfnamefont{F.~X.}
  \bibnamefont{Schmid}}, \bibinfo{journal}{Biochemistry}
  \textbf{\bibinfo{volume}{38}}, \bibinfo{pages}{2882} (\bibinfo{year}{1999}).

\bibitem[{\citenamefont{Mayor et~al.}(2000)\citenamefont{Mayor, Johnson,
  Daggett, and Fersht}}]{mayor00}
\bibinfo{author}{\bibfnamefont{U.}~\bibnamefont{Mayor}},
  \bibinfo{author}{\bibfnamefont{C.~M.} \bibnamefont{Johnson}},
  \bibinfo{author}{\bibfnamefont{V.}~\bibnamefont{Daggett}}, \bibnamefont{and}
  \bibinfo{author}{\bibfnamefont{A.~R.} \bibnamefont{Fersht}},
  \bibinfo{journal}{Proc. Natl. Acad. Sci. USA} \textbf{\bibinfo{volume}{97}},
  \bibinfo{pages}{13518} (\bibinfo{year}{2000}).

\bibitem[{\citenamefont{Crane et~al.}(2000)\citenamefont{Crane, Koepf, Kelly,
  and Gruebele}}]{crane00}
\bibinfo{author}{\bibfnamefont{J.~C.} \bibnamefont{Crane}},
  \bibinfo{author}{\bibfnamefont{E.~K.} \bibnamefont{Koepf}},
  \bibinfo{author}{\bibfnamefont{J.~W.} \bibnamefont{Kelly}}, \bibnamefont{and}
  \bibinfo{author}{\bibfnamefont{M.}~\bibnamefont{Gruebele}},
  \bibinfo{journal}{J. Mol. Biol.} \textbf{\bibinfo{volume}{298}},
  \bibinfo{pages}{283} (\bibinfo{year}{2000}).

\bibitem[{\citenamefont{Myers and Oas}(2002)}]{myers02}
\bibinfo{author}{\bibfnamefont{J.~K.} \bibnamefont{Myers}} \bibnamefont{and}
  \bibinfo{author}{\bibfnamefont{T.~G.} \bibnamefont{Oas}},
  \bibinfo{journal}{Annu. Rec. Biochem. and referenced there in.}
  \textbf{\bibinfo{volume}{71}}, \bibinfo{pages}{783} (\bibinfo{year}{2002}).

\bibitem[{\citenamefont{Hughes and Wang}(2003)}]{hughes03}
\bibinfo{author}{\bibfnamefont{W.~L.} \bibnamefont{Hughes}} \bibnamefont{and}
  \bibinfo{author}{\bibfnamefont{Z.~L.} \bibnamefont{Wang}},
  \bibinfo{journal}{Appl. Phys. Lett.} \textbf{\bibinfo{volume}{82}},
  \bibinfo{pages}{2886} (\bibinfo{year}{2003}).

\bibitem[{\citenamefont{Bargatin et~al.}(2005)\citenamefont{Bargatin, Myers,
  Arlett, Gudlewski, and Roukes}}]{bargatin05}
\bibinfo{author}{\bibfnamefont{I.}~\bibnamefont{Bargatin}},
  \bibinfo{author}{\bibfnamefont{E.~D.} \bibnamefont{Myers}},
  \bibinfo{author}{\bibfnamefont{J.}~\bibnamefont{Arlett}},
  \bibinfo{author}{\bibfnamefont{B.}~\bibnamefont{Gudlewski}},
  \bibnamefont{and} \bibinfo{author}{\bibfnamefont{M.~L.}
  \bibnamefont{Roukes}}, \bibinfo{journal}{Appl. Phys. Lett.}
  \textbf{\bibinfo{volume}{86}}, \bibinfo{pages}{133109}
  (\bibinfo{year}{2005}).

\bibitem[{\citenamefont{Hatfield and Quake}(1999)}]{hatfield99}
\bibinfo{author}{\bibfnamefont{J.~W.} \bibnamefont{Hatfield}} \bibnamefont{and}
  \bibinfo{author}{\bibfnamefont{S.~R.} \bibnamefont{Quake}},
  \bibinfo{journal}{Phys. Rev. Lett} \textbf{\bibinfo{volume}{82}},
  \bibinfo{pages}{3548} (\bibinfo{year}{1999}).

\bibitem[{\citenamefont{Patlak}(1993)}]{patlak93}
\bibinfo{author}{\bibfnamefont{J.~B.} \bibnamefont{Patlak}},
  \bibinfo{journal}{Biophys. J.} \textbf{\bibinfo{volume}{65}},
  \bibinfo{pages}{29} (\bibinfo{year}{1993}).

\bibitem[{\citenamefont{van Gunsteren and Berendsen}(1982)}]{vangunsteren82}
\bibinfo{author}{\bibfnamefont{W.~F.} \bibnamefont{van Gunsteren}}
  \bibnamefont{and} \bibinfo{author}{\bibfnamefont{H.~J.~C.}
  \bibnamefont{Berendsen}}, \bibinfo{journal}{Mol. Phys.}
  \textbf{\bibinfo{volume}{45}}, \bibinfo{pages}{637} (\bibinfo{year}{1982}).

\bibitem[{\citenamefont{Bourret et~al.}(1973)\citenamefont{Bourret, Frisch, and
  Pouquet}}]{bourret73}
\bibinfo{author}{\bibfnamefont{R.}~\bibnamefont{Bourret}},
  \bibinfo{author}{\bibfnamefont{U.}~\bibnamefont{Frisch}}, \bibnamefont{and}
  \bibinfo{author}{\bibfnamefont{A.}~\bibnamefont{Pouquet}},
  \bibinfo{journal}{Physica A} \textbf{\bibinfo{volume}{65}},
  \bibinfo{pages}{303} (\bibinfo{year}{1973}).

\bibitem[{\citenamefont{Brissaud and Frisch}(1974)}]{brissaud74}
\bibinfo{author}{\bibfnamefont{A.}~\bibnamefont{Brissaud}} \bibnamefont{and}
  \bibinfo{author}{\bibfnamefont{U.}~\bibnamefont{Frisch}},
  \bibinfo{journal}{J. Math. Phys.} \textbf{\bibinfo{volume}{15}},
  \bibinfo{pages}{524} (\bibinfo{year}{1974}).

\bibitem[{\citenamefont{van Kampen}(1975)}]{vankampen75}
\bibinfo{author}{\bibfnamefont{N.~G.} \bibnamefont{van Kampen}},
  \bibinfo{journal}{Phys. Rep.} p. \bibinfo{pages}{172} (\bibinfo{year}{1975}).

\bibitem[{\citenamefont{Roerdink}(1981)}]{roerdink81}
\bibinfo{author}{\bibfnamefont{J.~B. T.~M.} \bibnamefont{Roerdink}},
  \bibinfo{journal}{Physica A} \textbf{\bibinfo{volume}{109}},
  \bibinfo{pages}{23} (\bibinfo{year}{1981}).

\bibitem[{\citenamefont{Lindenberg et~al.}(1981)\citenamefont{Lindenberg,
  Seshadri, and West}}]{lindenberg81}
\bibinfo{author}{\bibfnamefont{K.}~\bibnamefont{Lindenberg}},
  \bibinfo{author}{\bibfnamefont{V.}~\bibnamefont{Seshadri}}, \bibnamefont{and}
  \bibinfo{author}{\bibfnamefont{B.~J.} \bibnamefont{West}},
  \bibinfo{journal}{Physica A} \textbf{\bibinfo{volume}{105}},
  \bibinfo{pages}{445} (\bibinfo{year}{1981}).

\bibitem[{\citenamefont{Roerdink}(1982)}]{roerdink82}
\bibinfo{author}{\bibfnamefont{J.~B. T.~M.} \bibnamefont{Roerdink}},
  \bibinfo{journal}{Physica A} \textbf{\bibinfo{volume}{112}},
  \bibinfo{pages}{557} (\bibinfo{year}{1982}).

\bibitem[{\citenamefont{Ikeda}(2000)}]{ikeda00}
\bibinfo{author}{\bibfnamefont{N.}~\bibnamefont{Ikeda}}, \bibinfo{journal}{J.
  Phys. A} \textbf{\bibinfo{volume}{33}}, \bibinfo{pages}{6385}
  (\bibinfo{year}{2000}).

\bibitem[{\citenamefont{Gradshteyn and Ryzhik}(1994)}]{intserprod}
\bibinfo{author}{\bibfnamefont{I.~S.} \bibnamefont{Gradshteyn}}
  \bibnamefont{and} \bibinfo{author}{\bibfnamefont{I.~M.}
  \bibnamefont{Ryzhik}}, \emph{\bibinfo{title}{Table of Integrals, Series, and
  Products}} (\bibinfo{publisher}{Academic Press}, \bibinfo{year}{1994}),
  \bibinfo{edition}{5th} ed.

\end{thebibliography}

\end{document}